\documentclass[11pt]{article}
\makeatletter
\usepackage[a4paper,top=28mm,bottom=30mm,left=27mm,right=27mm,
                headheight=14pt,headsep=8mm,footskip=11mm]{geometry}
\usepackage{xcolor}
\usepackage{fancyhdr}
\usepackage{titlesec}

\definecolor{arxivrule}{HTML}{0072B2}
\definecolor{arxivgrey}{gray}{0.35}

\newcommand{\shorttitle}[1]{\gdef\arxiv@shorttitle{#1}}
\newcommand{\affil}[1]{\gdef\arxiv@affil{#1}}
\newcommand{\email}[1]{\gdef\arxiv@email{#1}}
\gdef\arxiv@shorttitle{}
\gdef\arxiv@affil{}
\gdef\arxiv@email{}

\renewcommand{\maketitle}{%
  \thispagestyle{arxivfirst}%
  \begingroup
  \centering
  {\color{arxivrule}\rule{\textwidth}{1.6pt}\par}
  \vspace{5mm}
  {\LARGE\bfseries\@title\par}
  \vspace{4.5mm}
  {\color{arxivrule}\rule{\textwidth}{0.5pt}\par}
  \vspace{7mm}
  {\large\@author\par}
  \ifx\arxiv@affil\@empty\else
    \vspace{1.5mm}{\normalsize\itshape\arxiv@affil\par}\fi
  \ifx\arxiv@email\@empty\else
    \vspace{1mm}{\small\ttfamily\arxiv@email\par}\fi
  \vspace{4mm}
  {\small\color{arxivgrey}\@date\par}
  \vspace{7mm}
  \endgroup
  \let\arxiv@savedtitle\@title
}

\fancypagestyle{arxivfirst}{%
  \fancyhf{}%
  \fancyfoot[C]{\small\thepage}}
\renewenvironment{abstract}{%
  \small
  \list{}{\leftmargin=9mm\rightmargin=9mm\topsep=0pt\partopsep=0pt}%
  \item[]{\centering\bfseries\scshape Abstract\par}\vspace{1.5mm}%
  \noindent\ignorespaces}%
  {\endlist\vspace{4mm}}
\newcommand{\keywords}[1]{\par\medskip\noindent\textbf{Keywords:} #1}

\titleformat{\section}{\Large\bfseries}{\thesection}{0.8em}{}
\titleformat{\subsection}{\large\bfseries}{\thesubsection}{0.7em}{}
\titlespacing*{\section}{0pt}{5.5mm}{2.5mm}
\titlespacing*{\subsection}{0pt}{4.5mm}{2mm}
\makeatother
\usepackage[T1]{fontenc}
\usepackage[utf8]{inputenc}
\ifdefined\pdfgentounicode
  \input{glyphtounicode}
  \pdfglyphtounicode{ff}{0066 0066}
  \pdfglyphtounicode{fi}{0066 0069}
  \pdfglyphtounicode{fl}{0066 006C}
  \pdfglyphtounicode{ffi}{0066 0066 0069}
  \pdfglyphtounicode{ffl}{0066 0066 006C}
\fi
\usepackage{lmodern,microtype}
\DisableLigatures[f]{encoding = *, family = *}
\usepackage{amsmath,amssymb,amsthm,mathtools}
\usepackage{booktabs,tabularx,enumitem}
\usepackage{colortbl}
\usepackage{algorithm,algpseudocode}
\usepackage{tikz}
\usetikzlibrary{arrows.meta,decorations.pathreplacing,calc}
\definecolor{cbBlue}{HTML}{0072B2}
\definecolor{cbOrange}{HTML}{D55E00}
\definecolor{cbGreen}{HTML}{009E73}
\usepackage[colorlinks=true,linkcolor=blue!50!black,citecolor=blue!50!black,urlcolor=blue!50!black]{hyperref}
\usepackage{aliascnt}
\usepackage[capitalise,nameinlink,noabbrev]{cleveref}

\newtheorem{theorem}{Theorem}[section]
\newaliascnt{lemma}{theorem}
\newtheorem{lemma}[lemma]{Lemma}
\aliascntresetthe{lemma}
\newaliascnt{proposition}{theorem}
\newtheorem{proposition}[proposition]{Proposition}
\aliascntresetthe{proposition}
\newaliascnt{corollary}{theorem}
\newtheorem{corollary}[corollary]{Corollary}
\aliascntresetthe{corollary}
\theoremstyle{remark}
\newaliascnt{remark}{theorem}
\newtheorem{remark}[remark]{Remark}
\aliascntresetthe{remark}

\newcommand{\Ft}{\mathbb F_2}

\newcommand{\len}[1]{|#1|}

\newcommand{\cF}{\mathcal F}
\newcommand{\sg}{\sigma}
\newcommand{\Transition}{\textsc{Transition}}
\newcommand{\Trials}{\textsc{Trials}}
\newcommand{\Tape}{\textsc{Tape}}
\newcommand{\Windows}{\textsc{Windows}}
\algrenewcommand\algorithmicrequire{\textbf{Input:}}
\setlist{topsep=4pt,itemsep=2pt,parsep=0pt}
\numberwithin{equation}{section}

\title{Almost Optimal Deterministic Rendezvous\\ in Anonymous Graphs}
\shorttitle{Almost Optimal Deterministic Rendezvous in Anonymous Graphs}
\author{Debasish Pattanayak}
\affil{Indian Institute of Technology Indore, India}
\date{November 2026}

\begin{document}
\maketitle

\begin{abstract}
Two mobile agents, starting from different nodes of an unknown network, have
to meet at a node.  The network is modelled as a connected graph with $n$
nodes.  The nodes are unlabelled, and the ports at each node are numbered
locally.  The agents have distinct positive integer labels, move in
synchronous rounds, have unlimited memory and computational power, and learn
the port by which they enter a node.  What is the minimum time of
deterministic rendezvous in arbitrary graphs?  The best published lower bound,
$\Omega(n^2)$, is due to Dessmark et al.~[2006].  The best published upper
bound, $\widetilde O(n^5b)$, where $b$ is the length of the shorter label, is
due to Ta-Shma and Zwick~[2014] and was announced in 2007.  We close the gap
between the exponents $2$ and $5$ up to a logarithmic factor.  We first show
that, for every $n\ge12$, any algorithm and any two distinct labels, there is
a graph with $n$ nodes on which rendezvous takes at least $(1-o(1))\,n^4/1024$
rounds, and one of the agents has to traverse the same number of edges.  The
lower bound holds even if the agents start simultaneously and know $n$ and
both labels.  A bound of the same order holds on graphs in which the diameter
and the initial distance between the agents are both seven.  On the other
hand, we give an algorithm with which the agents meet within $O(n^4\log
n+n^2b)$ rounds after the start of the later agent.  Its agents know nothing
about the graph or about the other label, and they start with an arbitrary
delay.  Hence, for $b=O(n^2)$, the two bounds differ only by a factor $O(\log
n)$.  We also prove that, for every even $n$ and every $b\ge3$, some two
labels of length $b$ need $\Omega(n^2b/\log n)$ rounds.  The lower bound
carries over to gathering algorithms that also gather teams of two agents, to
the cost of asynchronous rendezvous, and to one agent that has to explore the
graph.

\end{abstract}

\section{Introduction}

\subsection{The background}

Rendezvous is one of the fundamental tasks performed by mobile agents in
networks.  Two agents start from different nodes of a network and have to
meet.  Depending on the application, the agents may be software agents that
navigate in a communication network, or mobile robots that move in a
system of corridors.  Software agents meet in order to exchange the data
that they collected at the nodes.  Robots meet in order to exchange samples
or to plan further actions.  In both cases the agents usually have no map
of the network, and the nodes may not show any identifiers.  It is thus
natural to ask how fast two agents can meet in an unknown network whose
nodes are anonymous.  The task has been extensively studied
(cf.\ the survey~\cite{Pelc23}).

Rendezvous is also the core of several other tasks.  In \emph{gathering}, a
team of more than two agents has to meet at one node, possibly although
some agents of the team are faulty~\cite{DPP14,Pelc18}.  Rendezvous is
gathering of a team of two agents, and gathering algorithms use rendezvous
algorithms as subroutines.  In \emph{treasure hunt}, one agent has to find
a target that does not move, called the treasure.  In \emph{exploration},
one agent has to visit all nodes.  An agent that explores the network finds
every treasure, and an agent that finds a treasure meets an agent that
stays at its starting node.  Hence a bound for one of these tasks bears on
the others.

In spite of this central role, the time of rendezvous in arbitrary networks
is not known.  For networks with $n$ nodes, the best lower bound has been
of order $n^2$ since the first paper on the subject~\cite{DFKP06}, and the
best algorithm, announced in 2007, takes about $n^5$
rounds~\cite{TZ14}.  Thus the gap between the exponents $2$ and $5$ has
been open for almost twenty years.  The survey of Pelc~\cite{Pelc23} lists
an optimal algorithm as an open problem.  The atlas of open problems
\emph{Mobility Matters}~\cite{Pat26} records the same gap, between
$\Omega(n^2+b)$ and $\widetilde O(n^5b)$.  Here the agents have distinct
integer labels, $b$ is the length of the shorter label, and the notation
$\widetilde O$ hides factors polylogarithmic in $n$ and $b$.  In this
paper we show that the right order is $n^4$, up to a logarithmic factor,
for labels of length $O(n^2\log n)$.

\subsection{The model and the problem}

The network is modelled as a finite simple undirected connected graph.  We
denote by $n$ the number of its nodes, which we call the \emph{size} of the
graph, and by $m$ the number of its edges.  We use the standard notions of
graph theory.  Two nodes are \emph{adjacent}, and each is a \emph{neighbour}
of the other, if an edge joins them.  The \emph{degree} of a node is the
number of its neighbours.  A graph is \emph{$d$-regular} if all its nodes
have degree $d$.  The \emph{length} of a path is the number of its edges.
The \emph{distance} between two nodes is the length of a shortest path
between them, and the \emph{diameter} of a graph is the largest distance
between two of its nodes.  The \emph{clique} $K_q$ is the graph with $q$
nodes in which every two nodes are adjacent.

Nodes are unlabelled.  Such a
graph is called \emph{anonymous}.  The ports at any node of degree $d$ are
arbitrarily numbered $1,\dots,d$.  The assignment of the port numbers at
all nodes is called the \emph{port labelling}, and a graph with a port
labelling is called a \emph{port-labelled graph}.  There is no relation
between the two port numbers of an edge.  Nodes cannot be marked.

Time is discretized into rounds.  In each round an agent either stays at
its current node or takes one of the ports at this node.  An agent that
takes a port is at the other end of the corresponding edge at the end of
the round.  We say that the agent \emph{traverses} the edge.  When an agent
enters a node, it learns the degree of the node and the number of the port
by which it enters.  These two numbers are the \emph{observations} of the
agent in the round.  At its start, an agent observes only the degree of its
node.  Agents have unlimited memory and are computationally
unbounded.  Hence, an agent remembers everything that it has seen, and
local computations take no time.

Two agents \emph{meet} when they are at the same node at the end of a
round.  The task of \emph{rendezvous} is to make the two agents meet.  Agents that traverse the same edge in opposite directions in the
same round do not notice this fact.  Such agents do not meet.
Prior to the meeting, an agent has no information about the other agent.

Each agent is placed at a node, called its \emph{starting node}, in the
round in which it starts.  The two starting nodes are different.  An
adversary chooses the graph, its port numbers, the starting nodes of the
agents and the round in which each agent starts.  The agents start
\emph{simultaneously} if they start in the same round.  Otherwise the
number of rounds between the two starts is called the \emph{delay}.  An
agent does not exist in the graph before its start.

Agents that are identical cannot always meet.  Consider a ring in which,
at every node, port $1$ leads clockwise and port $2$ leads
counterclockwise.  Two identical agents that start simultaneously in such a
ring take the same ports in every round, so the distance between them never
changes.  In order to avoid this impossibility, each agent has a
\emph{label}, which is a positive integer.  The two labels are different.
Both agents execute the same deterministic algorithm, and the label of an
agent is a parameter of the algorithm.  An agent knows its own label.  We
say that the agents \emph{know} some other information, such as $n$, if
this information is given to both agents as an input of the algorithm.
Unless we say otherwise, the agents know nothing about the graph or about
the other label.  In every round, the algorithm determines from the label of an
agent, its input and its observations in the previous rounds whether the
agent stays or which port it takes.  A \emph{rendezvous algorithm} is an algorithm with which the
agents meet in every graph, from all starting nodes and for all labels.  We denote the labels by $\ell_1$
and $\ell_2$.  The \emph{length} of a label $\ell$ is the number
$\len{\ell}=\lfloor\log\ell\rfloor+1$ of bits of its binary representation.
Throughout the paper, $b=\min\{\len{\ell_1},\len{\ell_2}\}$ denotes the
length of the shorter label, and all logarithms are to base $2$.

The \emph{time} of a
rendezvous algorithm is the number of rounds from the start of the later
agent until the meeting.  The \emph{cost} of an agent is the number of
edges that it traverses until the meeting.  The \emph{initial distance} is
the distance between the starting nodes.

In this paper we study the following problem.  What is the minimum time of
deterministic rendezvous, in the worst case over all graphs of size $n$?

\subsection{Our results}

Our first result is a lower bound.  The best previously published lower
bound for arbitrary graphs is $\Omega(n^2)$, due to Dessmark et
al.~\cite{DFKP06}.  We raise the exponent from $2$ to $4$.

\begin{theorem}[Lower bound]
\label{thm:quartic}
Let $n\ge12$ and $q=\lfloor n/4\rfloor$, and let
\[
 L(n)=\binom q2\binom{q-1}2=\frac{q(q-1)^2(q-2)}4 .
\]
For any deterministic rendezvous algorithm and any two distinct labels,
there exist a graph of size $n$ and two starting nodes such that, if the
agents start simultaneously, then
\begin{enumerate}[label=(\alph*)]
\item the agents do not meet during the first $L(n)-1$ rounds, and
\item at least one of the agents traverses at least $L(n)$ edges until the
meeting.
\end{enumerate}
The graph has diameter at most $7$.  It is $(q-1)$-regular if $n$
is divisible by $4$.  The statement holds even if both agents know $n$ and
both labels, and even if the two agents execute different algorithms.
\end{theorem}

Since $L(n)=n^4/1024-O(n^3)$, both the time and the cost of rendezvous are
at least $(1-o(1))\,n^4/1024$ in the worst case.  The agents of
\cref{thm:quartic} have every advantage that the models of
\cref{tab:literature} give to agents.  They start simultaneously, know the
size of the graph, learn the ports by which they enter, and have unlimited
memory.  Hence the lower bound holds in each of these models.

A small diameter and a small initial distance do not help the agents.  We
show a lower bound of the same order of magnitude on graphs in which the
diameter and the initial distance are both equal to seven
(\cref{thm:seven}).  For sparse graphs, the argument yields the lower bound
$\Omega(m^2)$ on regular graphs with $m$ edges (\cref{rem:edges}).

The bound of \cref{thm:quartic} does not depend on the labels.  We also
prove a lower bound that grows with the length of the labels.

\begin{theorem}[Lower bound in terms of the labels]
\label{thm:labels}
Let $n\ge6$ be even, let $q=n/2$ and $s=\binom q2$.  Let $\Lambda$ be a set
of $L\ge2$ labels, and let
$k=\lfloor(\log L-1)/\log(\mathrm es)\rfloor$.  For any deterministic
rendezvous algorithm, there exist two distinct labels of $\Lambda$, a graph
of size $n$ and two starting nodes such that agents with these labels that
start simultaneously do not meet during the first
$\lceil(k+1)s/2\rceil-1$ rounds.  The diameter of the graph and the initial
distance are both equal to $3$.  The statement holds even if both agents
know $n$ and $\Lambda$.
\end{theorem}

For the set of all labels of length $b$, the bound of \cref{thm:labels} is
at least of order $n^2b/\log n$ (\cref{cor:labels}).  Hence, for even $n$, in the
worst case over the graphs of size $n$ and the labels of length $b$, the
time of rendezvous is $\Omega(n^4+n^2b/\log n)$.  The same holds for odd $n$
that are divisible by $3$ (\cref{rem:three}).

Our second result is an algorithm whose time almost matches the lower
bounds.  Its agents need no information.

\begin{theorem}[Upper bound]
\label{thm:upper}
There exists a deterministic algorithm with the following property.  Two
agents that have distinct labels and know nothing about the graph or about
the other label meet within $O(n^4\log n+n^2b)$ rounds after the start of
the later agent.  The bound holds in any connected graph of size $n$ and for
any delay.
\end{theorem}

By \cref{thm:quartic}, the time of the algorithm is optimal up to the factor
$O(\log n)$ whenever $b=O(n^2\log n)$.  By \cref{thm:labels}, for even $n$, the term
$n^2b$ cannot be improved by more than the factor $\log n$ by an algorithm
whose agents do not know the label of the other agent.  The agents of the
lower bounds start simultaneously and know $n$.  The agents of
\cref{thm:upper} know nothing and start with an arbitrary delay.  Hence a
simultaneous start and the knowledge of $n$ change the time of rendezvous at
most by a logarithmic factor.  On the other hand, the knowledge of the other label removes the term
$n^2b$.  Indeed, suppose that both agents know $n$ and both labels.  Then the
agent with the smaller label stays at its node, and the other agent repeats
the exploration of \cref{rem:explore} for $N=n$.  The first exploration that begins after
the start of the later agent visits the node of the other agent.  Hence the
agents meet within
$O(n^4\log n)$ rounds after the start of the later agent, for any delay and
for labels of any length.  Thus the term $n^2b$ is the price of not knowing
the other label.  However, the bound of \cref{thm:quartic} holds even for
agents that know both labels.

The best published algorithm, due to Ta-Shma
and Zwick~\cite{TZ14}, works in time $\widetilde O(n^5b)$.
Compared with this algorithm, we lower the exponent of $n$ from $5$ to $4$,
and the length of the label becomes an additive term.  On the other hand,
the agents of Ta-Shma and Zwick do not learn the ports by which they enter.
Like ours, their algorithm uses sequences whose existence is proved by the
probabilistic method, and its local computations are not known to take
polynomial time.  Ta-Shma and Zwick also give an algorithm whose local
computations take polynomial time, for agents that learn the ports by which
they enter.  Its time is a polynomial of larger degree.  Our agents return by the
ports by which they enter, and they find what to do
by an exhaustive search.  The model permits such a search, because agents are
computationally unbounded.

We do not know whether the factor $\log n$ is necessary.  We show that it
is not necessary on a family of graphs on which the lower bound
$\Omega(n^4)$ still holds.  These graphs are built from a clique $K_q$ by
taking four copies of each node, and they are called four-sheet covers of
$K_q$.  We define covers and the cyclic port labelling of $K_q$ in
\cref{sec:prelim}.

\begin{corollary}[Four-sheet covers of a clique]
\label{cor:fourcover}
Let $q\ge3$ and $n=4q$, and suppose that both agents know $q$.  On the connected
four-sheet covers of the clique $K_q$ with the cyclic port labelling, the
minimum time of rendezvous is $O(n^4+n^2b)$.  It is $\Theta(n^4)$ whenever
$b=O(n^2)$.  The upper bound holds for any delay.
\end{corollary}

Finally, we explain why our methods do not remove the factor $\log n$
(\cref{sec:log}).  \Cref{tab:literature} compares our results with the
published bounds.

\begin{table}[ht]
\centering\small
\renewcommand{\arraystretch}{1.3}
\begin{tabularx}{\linewidth}{>{\raggedright\arraybackslash}p{.36\linewidth}p{.13\linewidth}>{\raggedright\arraybackslash}Xp{.13\linewidth}}
\rowcolor{black!75}
\color{white}\textbf{Time} & \color{white}\textbf{Start} &
 \color{white}\textbf{Further assumptions} & \color{white}\textbf{Source}\\
\rowcolor{cbOrange!30}
\multicolumn{4}{l}{\textbf{Lower bounds}}\\
$\Omega(n^2)$ & simultaneous & some family of graphs; $n$ known &
 \cite{DFKP06}, 2006\\
$\Omega(nb)$ & simultaneous & rings; some two labels of length at most $b$ & \cite{DFKP06}, 2006\\
\rowcolor{cbOrange!12}
$(1-o(1))\,n^4/1024$ & simultaneous & every $n\ge12$; any two labels; $n$
 and both labels known; also a bound on the cost & \Cref{thm:quartic}\\
\rowcolor{cbOrange!12}
$\Omega(n^2b/\log n)$ & simultaneous & even $n\ge6$; some two labels of
 length $b\ge3$; $n$ known & \Cref{thm:labels}\\
\rowcolor{cbBlue!30}
\multicolumn{4}{l}{\textbf{Upper bounds}}\\
$O(n^5\sqrt{\tau\log\ell}\log n+n^{10}\log^2n\log\ell)$ & delay $\tau$ &
 smaller label $\ell$ & \cite{DFKP06}, 2006\\
$O(\log^3\ell+n^{15}\log^{12}n)$ & any delay & smaller label $\ell$; agents
 learn the ports by which they enter & \cite{KM08}, 2008\\
$\widetilde O(n^5b)$ & any delay & $n$ unknown; agents do not learn the
 ports by which they enter & \cite{TZ14}, 2014\\
\rowcolor{cbBlue!12}
$O(n^4\log n+n^2b)$ & any delay & $n$ unknown; agents learn the ports by
 which they enter; exhaustive search & \Cref{thm:upper}\\
\bottomrule
\end{tabularx}
\caption{Time of deterministic rendezvous of two labelled agents in arbitrary
graphs of size $n$.  The length of the shorter label is $b$.  The shaded
rows are the bounds of this paper.  The other rows are earlier bounds, with the year of the
journal version.  See also the survey~\cite{Pelc23} and the
atlas~\cite{Pat26}.  The algorithm of~\cite{TZ14} was announced in 2007.  The
exponent of $n$ is $4$ in both directions.  The term $n^4\log n$ of the
upper bound is tight up to a factor $O(\log n)$.  So is the term $n^2b$,
for even $n$ and agents that do not know the other label.}
\label{tab:literature}
\end{table}

\subsection{Consequences for related problems}
\label{sec:consequences}

Our bounds have consequences beyond rendezvous of two agents in the model
of this paper.  We list them, and we point to the place where each of them
is proved.

\emph{The open problem of rendezvous.}  Before this paper, the time of
rendezvous in arbitrary graphs was known to be between $\Omega(n^2)$ and
$\widetilde O(n^5b)$.  \Cref{thm:quartic,thm:upper} place it between
$\Omega(n^4)$ and $O(n^4\log n+n^2b)$.  Hence the exponent of $n$ is $4$.
For the labels,
Dessmark et al.~\cite{DFKP06} proved that rendezvous takes $\Omega(b)$
rounds, and $\Omega(nb)$ rounds on rings.  \Cref{thm:labels} raises this
bound to the order $n^2b/\log n$, for even $n$, at initial distance $3$.
Thus, for even $n$ and for $n$ divisible by $3$, the time is
$\Omega(n^4+n^2b/\log n)$, and each of the two terms of
the upper bound is tight up to a factor $O(\log n)$.  The lower bounds hold
for agents that know $n$ and start simultaneously, and the upper bound
holds for agents that know nothing and start with an arbitrary delay.
Agents that know both labels can avoid the term $n^2b$ (see the discussion
after \cref{thm:upper}).

\emph{Small initial distance.}  For agents that start simultaneously,
Dessmark et al.~\cite{DFKP06} gave an algorithm whose time depends on the
initial distance and on the maximum degree of the graph, and is exponential
in the initial distance.  It is very fast when both parameters are small.
\Cref{thm:seven} shows that a small initial distance alone does not help.
At initial distance seven, rendezvous still takes $\Omega(n^4)$ rounds on
graphs of large degree.  At initial distance three, some two labels of
length $b$ need $\Omega(n^2b/\log n)$ rounds (\cref{thm:labels}).

\emph{One agent: treasure hunt and exploration.}  The argument of the lower
bound, applied to one agent, shows that an agent needs
$(1-o(1))\,n^4/64$ rounds to explore the graph, and hence to find a
treasure in the worst case, even if it knows $n$ (\cref{cor:treasure}).  For
treasure hunt by an agent that learns the port by which it enters, the
atlas~\cite{Pat26} records no lower bound in terms of $n$.  On the other hand, an
agent that knows an upper bound $N$ on $n$ visits all nodes within
$O(N^4\log N)$ rounds (\cref{rem:explore}).  Hence exploration and
rendezvous have the same time up to logarithmic factors, when the labels
are short.    Kosowski~\cite{Kos14} reports an exploration of length
$O(N^4\log^2N)$ by a derandomized Metropolis walk, with reference
to~\cite{NOSY10}.  The explicit algorithms of~\cite{TZ14,Xin26} follow
universal exploration sequences.  Such a sequence tells the agent in every
round by how much the port that it takes differs from the port by which it
entered.  The length of the known explicit sequences is a polynomial of
large degree.

\emph{Gathering.}  An algorithm that gathers every team of at least two
agents solves rendezvous when the team has two agents.  Hence
\cref{thm:quartic} shows that such an algorithm needs $\Omega(n^4)$ rounds
on some graph (\cref{rem:models}).  The earlier lower bounds on the time
of gathering in arbitrary graphs are at most quadratic.  For a team of two
agents, the bound $\Omega(n^2)$ of Dessmark et al.~\cite{DFKP06} applies,
and Molla et al.~\cite{MMM23} note the bound $\Omega(n)$ for two groups of
agents at the ends of a path.  For an algorithm that gathers a team of $k$
agents, we show the lower bound $\Omega(n^4/k^4)$ (\cref{cor:teams}).
Conversely, a reduction of Kowalski and Malinowski~\cite{KM08} turns
Algorithm \Windows\ into an algorithm that gathers every team of agents with
distinct labels within $O(n^4\log n+n^2b)$ rounds, where $b$ is the length
of the smallest label (\cref{rem:gathering}).  Hence the time of gathering
every team is also determined up to a factor $O(\log n)$, when the labels
are short.  The lower bounds concern small teams.  Indeed, Molla et
al.~\cite{MMM23} gather a team of more than $n/2$ agents in
$O(n^3)$ rounds, and the agents detect that the gathering is complete, when the agents know $n$, start simultaneously and have
labels in $\{1,\dots,n^c\}$ for a constant $c$.  Thus a large team gathers faster than two agents can
meet.  In the other direction, the time bounds of gathering algorithms are
often multiples of the time of the rendezvous algorithm that they use as a
subroutine~\cite{DPP14}.  A faster rendezvous algorithm is thus the first
step towards faster gathering.

\emph{Asynchronous agents.}  In the asynchronous model, an adversary chooses
the speed of each agent, and the cost of an algorithm is the total number
of edge traversals.  Rendezvous at polynomial cost is
possible~\cite{DPV15}.  The adversary may let both agents move at the same
speed.  Hence part (b) of \cref{thm:quartic} shows that the cost of
asynchronous rendezvous is $\Omega(n^4)$ (\cref{rem:models}).

\emph{Weaker agents.}  As noted after \cref{thm:quartic}, its agents have every
advantage.  Hence the lower bound holds for agents with bounded memory and for agents
that do not learn the ports by which they
enter.  In the last model, which is the model of~\cite{TZ14}, the time of
rendezvous is thus between $\Omega(n^4)$ and $\widetilde O(n^5b)$.

\subsection{Related work}

Rendezvous of labelled agents in arbitrary graphs was first studied by
Dessmark et al.~\cite{DFKP06}.  They gave the first algorithm whose time is
polynomial in the size of the graph, in the length of the smaller label and
in the delay.  They also proved the lower bound $\Omega(n^2)$ on some family
of graphs.  Kowalski and Malinowski~\cite{KM08} removed the dependence of
the time on the delay.  Ta-Shma and Zwick~\cite{TZ14} obtained the time
$\widetilde O(n^5b)$ for agents that do not learn the ports by which they
enter.  The
survey of Pelc~\cite{Pelc23} reports $\Omega(n^2)$ as the lower bound for
arbitrary graphs and names the algorithm of~\cite{TZ14} as the most
efficient one.  It lists an optimal synchronous rendezvous algorithm for
arbitrary graphs as an open problem.  Pattanayak and Pelc~\cite{PP24}
study treasure hunt and rendezvous in graphs whose nodes may have infinite
degree, and ask for an optimal rendezvous algorithm even for finite graphs.

Our proofs use tools from the study of universal traversal sequences.  A
\emph{universal traversal sequence} for a family of graphs is a sequence of
port numbers with the following property.  An agent that takes the ports of
the sequence one after another visits all nodes, in every graph of the family
and from every starting node.  Aleliunas et al.~\cite{AKLLR79} proved that
short universal traversal sequences exist.  Their argument takes a random
sequence and bounds the number of graphs on which the sequence may fail.
Our upper bound follows the same argument.  Borodin et al.~\cite{BRT92}
proved that a universal traversal sequence for the $d$-regular graphs of
size $n$ has length $\Omega(d^2n^2+dn^2\log(n/d))$, for $3\le d\le
n/3-2$.  Their bound is $\Omega(n^4)$ for linear degrees $d\le n/3-2$.  For
$d\ge n/2$, universal traversal sequences of length $O(n^3\log n)$ exist,
as Borodin et al.\ report.  For a fixed sequence that is too short, the proof of Borodin et al.~\cite{BRT92} constructs a graph in
which the sequence never traverses one of the edges.  The graph is found by
solving linear equations.  The argument of our \cref{lem:hide} is theirs.

The bound of Borodin et al.~\cite{BRT92} concerns one agent that has to find a
stationary target.  It does not imply a bound for rendezvous.  Indeed,
neither of the two agents can be assumed to stay at its starting node, and two agents may
meet although neither of them has visited all nodes.  The new step of our
proof is to construct such a graph for each of the two moving agents
separately.  Then we join the two graphs through the edges that the agents
never traverse.  The idea of
separating two agents by edges that neither of them has traversed goes back
to the lower bound of Dessmark et al.~\cite{DFKP06}.

Recent papers on rendezvous concern other questions.
Das~\cite{Das26} studies the feasibility of rendezvous for anonymous agents
that leave footprints.  Das and Pelc~\cite{DP26} study rendezvous of finite
automata that can drop pebbles.  Xin~\cite{Xin26} improves the exploration
sequences on which the algorithm of~\cite{TZ14} is based, for the task of
one agent that has to find a stationary target.  The paper~\cite{Xin26} gives no new bound for rendezvous of two moving
agents.  To the best
of our knowledge, no lower bound above $\Omega(n^2)$ was published for the
time of rendezvous in arbitrary graphs.

The paper is organized as follows.  \Cref{sec:prelim} contains the notions
used in the proofs.  \Cref{sec:quartic} proves the lower bound, and
\cref{sec:seven} proves its variant for graphs of diameter seven.
\Cref{sec:upper} presents the algorithms and proves the upper bounds.
\Cref{sec:log} discusses the factor $\log n$, and \cref{sec:conclusion}
contains conclusions and open problems.

\section{Preliminaries}
\label{sec:prelim}

In this section, we define the notions used in the rest of the paper.  We
first describe the behaviour of an agent by a sequence of actions.  Next, we
fix the port labelling of the clique and define the covers of a graph.  We
conclude the section by a lemma, which says that on covers of the clique an
agent executes a fixed sequence of actions.

An \emph{action} is an integer.  The action $0$ means that the agent stays
at its current node for one round.  An action $p\ge1$ means that the agent
takes port $p$.  A \emph{word} is a finite sequence of actions.  The
\emph{moves} of a word are its nonzero actions.  Consider a port-labelled
graph, a node $x$ and a word $w$ whose actions are at most the degrees of
the nodes at which they are executed.  The \emph{walk of $w$ from $x$} is
the sequence of nodes visited by an agent that starts at $x$ and executes
the actions of $w$ in consecutive rounds.  The walk \emph{traverses} an edge
whenever the agent does.  We denote by $w[1..t]$ the word formed by the
first $t$ actions of $w$.

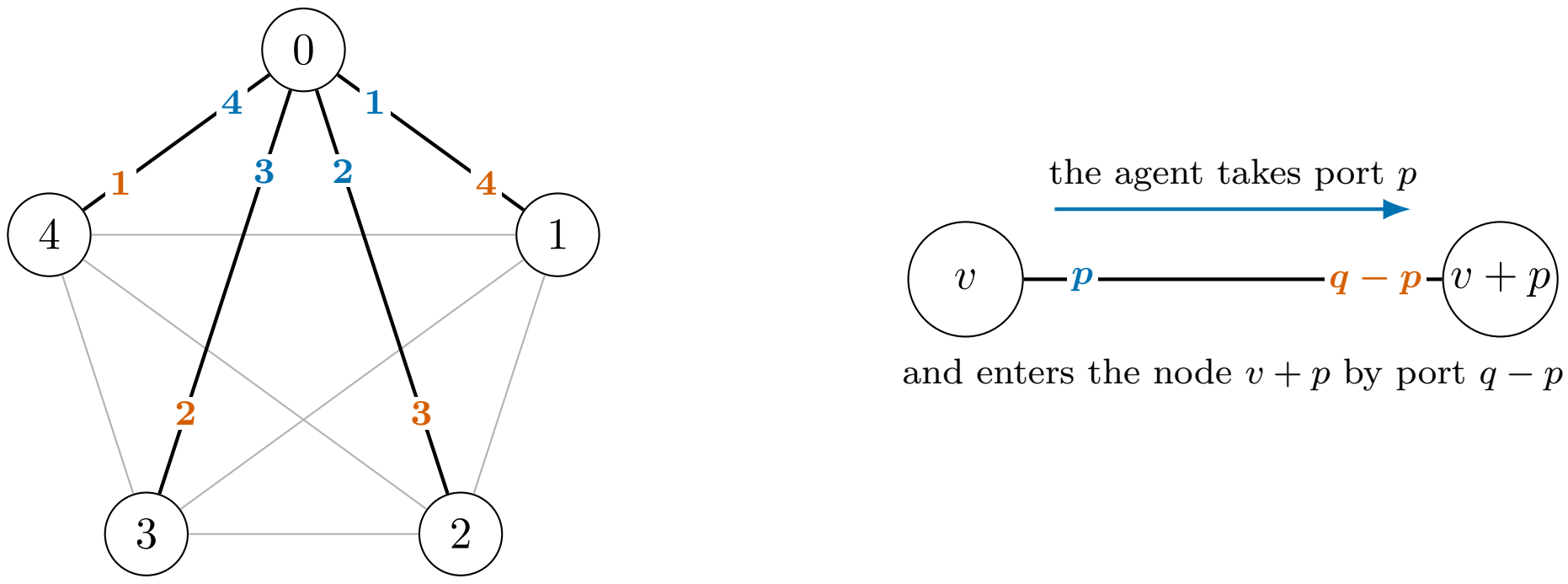
\begin{figure}[ht]
\centering
\begin{tikzpicture}[x=1cm,y=1cm,
  nd/.style={circle,draw,fill=white,inner sep=0.5pt,minimum size=6.5mm,font=\small},
  pl/.style={fill=white,inner sep=1pt,font=\scriptsize}]
\foreach \v in {0,...,4} \node[nd] (v\v) at ({90-72*\v}:2.1) {$\v$};
\foreach \a/\b in {1/2,1/3,1/4,2/3,2/4,3/4} \draw[black!30] (v\a)--(v\b);
\foreach \p/\qp in {1/4,2/3,3/2,4/1}
  \draw[thick] (v0)-- node[pl,pos=0.2,text=cbBlue]{\textbf{\p}}
                      node[pl,pos=0.8,text=cbOrange]{\textbf{\qp}} (v\p);
\node[font=\small] at (0,-2.75) {(a) the ports of the edges at node $0$ of $K_5$};
\begin{scope}[shift={(5.2,0.3)}]
\node[nd,minimum size=9mm] (a) at (0,0) {$v$};
\node[nd,minimum size=9mm] (b) at (4.2,0) {$v+p$};
\draw[thick] (a)-- node[pl,pos=0.14,text=cbBlue]{$\boldsymbol p$}
                   node[pl,pos=0.84,text=cbOrange]{$\boldsymbol{q-p}$} (b);
\draw[-{Latex[length=2.2mm]},cbBlue,thick] (0.7,0.55)-- node[above,font=\scriptsize,text=black]
   {the agent takes port $p$} (3.5,0.55);
\node[font=\scriptsize,align=center] at (2.1,-0.75)
   {and enters the node $v+p$ by port $q-p$};
\node[font=\small] at (2.1,-3.05) {(b) the rule for every edge of $K_q$};
\end{scope}
\end{tikzpicture}
\caption{The cyclic port labelling of the clique $K_q$, whose nodes are
$0,\dots,q-1$.  (a) The clique $K_5$.  At node $0$, port $p$ (blue) belongs to
the edge that leads to node $p$.  At the other end, the same edge has port
$5-p$ (orange).  The ports at the other nodes are numbered by the same rule.
(b) In general, port $p$ at node $v$ leads to node $v+p$ modulo $q$.  Hence an
agent that takes port $p$ always enters by port $q-p$, and the port by which
it enters gives it no information.}
\label{fig:ports}
\end{figure}

Throughout the paper, the nodes of the clique $K_q$ are the integers
$0,\dots,q-1$, and sums of nodes are computed modulo $q$.  Port
$p\in\{1,\dots,q-1\}$ at node $v$ leads to node $v+p$.  We call this port labelling the
\emph{cyclic port labelling} (\cref{fig:ports}).  An agent that takes port
$p$ enters the next node by port $q-p$.  We denote by $s=\binom q2$ the
number of edges of $K_q$.  Recall that a \emph{spanning tree} of a connected graph is a set of edges
that joins all nodes and contains no cycle.  Every spanning tree of $K_q$
has $q-1$ edges.  We
denote by $\beta=s-q+1=\binom{q-1}2$ the number of the remaining edges,
which is called the \emph{cycle rank} of $K_q$.

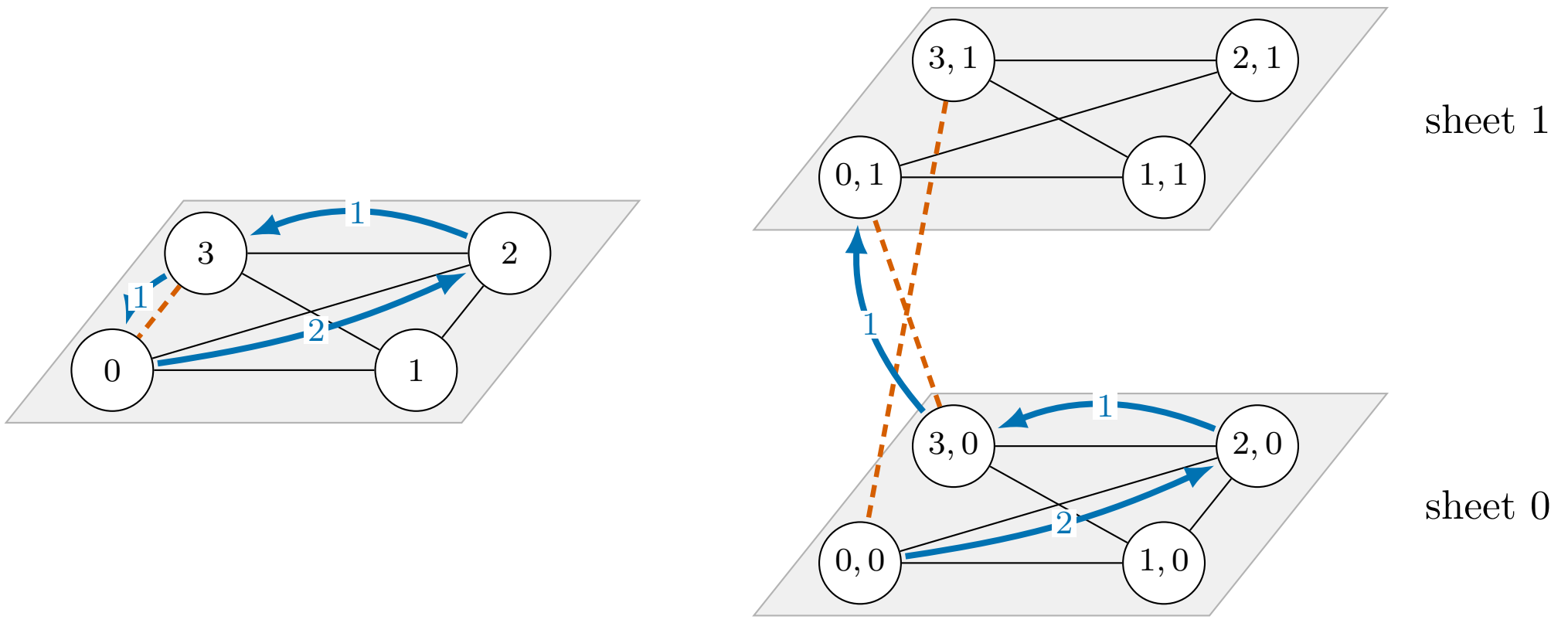
\begin{figure}[ht]
\centering
\begin{tikzpicture}[x=1cm,y=1cm,
  nd/.style={circle,draw,fill=white,inner sep=0.3pt,minimum size=7mm,font=\scriptsize},
  sg/.style={cbOrange,very thick,densely dashed},
  wk/.style={cbBlue,line width=1.5pt,-{Latex[length=2.6mm]},shorten >=1pt,shorten <=1pt},
  pl/.style={fill=white,inner sep=0.8pt,font=\scriptsize,text=cbBlue}]
\foreach \s/\ox/\oy in {b/0/1.65,0/6.4/0,1/6.4/3.3}{
  \fill[black!6,draw=black!30] (\ox-0.91,\oy-0.45)--(\ox+2.99,\oy-0.45)
     --(\ox+4.51,\oy+1.45)--(\ox+0.61,\oy+1.45)--cycle;
  \coordinate (c\s0) at (\ox,\oy);
  \coordinate (c\s1) at (\ox+2.6,\oy);
  \coordinate (c\s2) at (\ox+3.4,\oy+1);
  \coordinate (c\s3) at (\ox+0.8,\oy+1);
}
\foreach \v in {0,1,2,3}{
  \node[nd] (b\v) at (cb\v) {$\v$};
  \node[nd] (n\v0) at (c0\v) {$\v,0$};
  \node[nd] (n\v1) at (c1\v) {$\v,1$};
}
\foreach \a/\b in {0/1,1/2,2/3,1/3,0/2}{
  \draw (b\a)--(b\b); \draw (n\a0)--(n\b0); \draw (n\a1)--(n\b1);
}
\draw[sg] (b3)--(b0);
\draw[sg] (n30)--(n01);
\draw[sg] (n31)--(n00);
\draw[wk] (b0) to[bend right=8] node[pl,pos=0.5]{2} (b2);
\draw[wk] (b2) to[bend right=22] node[pl,pos=0.5]{1} (b3);
\draw[wk] (b3) to[bend right=22] node[pl,pos=0.5]{1} (b0);
\draw[wk] (n00) to[bend right=8] node[pl,pos=0.5]{2} (n20);
\draw[wk] (n20) to[bend right=22] node[pl,pos=0.5]{1} (n30);
\draw[wk] (n30) to[bend left=22] node[pl,pos=0.5]{1} (n01);
\node[font=\small,anchor=west] at (11.1,0.5) {sheet $0$};
\node[font=\small,anchor=west] at (11.1,3.8) {sheet $1$};
\node[font=\small] at (1.7,-1.0) {(a) the base $K_4$ with a signing $\sg$};
\node[font=\small] at (8.2,-1.0) {(b) the two-sheet cover $C_\sg$};
\draw[sg] (-0.6,-1.7)--(0.3,-1.7) node[right,black,font=\scriptsize]{edge with $\sg=1$, and its two copies};
\draw (5.3,-1.7)--(6.2,-1.7) node[right,font=\scriptsize]{edge with $\sg=0$, and its two copies};
\end{tikzpicture}
\caption{A two-sheet cover.  (a) The base is the clique $K_4$ with the cyclic
port labelling.  The signing $\sg$ has the value $1$ on the edge $\{0,3\}$
(dashed) and the value $0$ on the other five edges.  (b) The cover $C_\sg$
has two nodes $(v,0)$ and $(v,1)$ for every node $v$ of the base.  The nodes
$(v,0)$ form sheet $0$, and the nodes $(v,1)$ form sheet $1$.  Each node is
drawn above its projection.  An edge with $\sg=0$ has one copy inside each
sheet.  The two copies of the edge $\{0,3\}$ go from one sheet to the other.
Every node has the same ports as its projection.  The blue arrows show the
walk of the word $(2,1,1)$, with the port taken in each round.  In the base
the walk from node $0$ returns to node $0$.  In the cover the walk from
$(0,0)$ visits the same projections, but its last move traverses
a copy of $\{0,3\}$.  Hence it changes its sheet and ends at $(0,1)$, above
its starting node.  The agent observes
the same degrees and ports in both graphs.}
\label{fig:cover}
\end{figure}

We now define covers.  Informally, a cover of a graph $B$ consists of
several copies of the node set of $B$, which are called sheets.  Every edge
of $B$ is copied once for each sheet.  A copy either stays inside a sheet or
leads to another sheet.  An agent in the cover sees the same degrees and
ports as an agent in $B$.  Hence it cannot observe in which sheet it is.
\Cref{fig:cover} shows an example with two sheets.

Formally, let $B$ be a connected port-labelled graph, and let $k\ge2$ be an
integer.  A \emph{$k$-sheet cover} of $B$ is a graph defined as follows.
Its nodes are the pairs $(v,a)$, where $v$ is a node of $B$ and
$a\in\{0,\dots,k-1\}$.  The number $a$ is called the \emph{sheet} of the
node $(v,a)$.  We also call the set of all nodes $(v,a)$ with the same $a$
the sheet $a$.  Every edge $\{u,v\}$ of $B$, with a fixed order of its ends,
is assigned a permutation of the sheets.  The edge is replaced by $k$
edges.  For every sheet $a$, one of them joins the node $(u,a)$ to the node
$(v,a')$, where $a'$ is the image of $a$ under the permutation.  Each of
these edges is called a \emph{copy} of $\{u,v\}$.  A copy has the same port numbers as the edge
$\{u,v\}$, at the corresponding ends.  The graph $B$ is called the
\emph{base} of the cover.  The node $v$ is called the \emph{projection} of
the node $(v,a)$.  Hence a cover of $B$ has $k$ times as many nodes as $B$,
and every node of the cover has the same degree and the same port numbers
as its projection.  A cover need not be connected.

Covers are a standard notion of graph theory, in which they are also
called covering graphs or lifts.  Amit and Linial~\cite{AL02} define a
$k$-lift of a graph by one permutation for each edge, as above.
Angluin~\cite{Ang80} used covers to prove impossibility results for
anonymous networks of processors.  Bilu and Linial~\cite{BL06} describe the
two-sheet covers of a graph, which they call $2$-lifts, by signings.

Two-sheet covers are described by bits.  We denote by $\Ft$ the set
$\{0,1\}$ with addition and multiplication modulo $2$.  Thus $1+1=0$ in
$\Ft$.  For $k=2$ a permutation of the sheets either fixes both sheets or
exchanges them.  A two-sheet cover is thus described by a function $\sg$
that assigns a bit $\sg(e)\in\Ft$ to every edge $e$ of $B$.  We call such a
function a \emph{signing}.  For an edge $\{u,v\}$ we write $\sg(uv)$ for
$\sg(\{u,v\})$.  The two-sheet cover $C_\sg$ has the copies
$\{(u,a),(v,a+\sg(uv))\}$ of every edge $\{u,v\}$, for $a\in\Ft$, where the
sum is computed in $\Ft$.  An agent
that traverses a copy of an edge $e$ changes its sheet if and only if
$\sg(e)=1$.  Consider the walk of a word from a node $(x,a)$ of a cover.
The projections of its nodes form the walk of the same word from $x$ in the
base.  Only the sheets depend on the cover.

We use one fact of linear algebra.  For an integer $\beta\ge1$, we denote
by $\Ft^\beta$ the set of vectors with $\beta$ coordinates in $\Ft$.  The
\emph{scalar product} of two vectors $x,y\in\Ft^\beta$ is
$x\cdot y=x_1y_1+\dots+x_\beta y_\beta$, computed in $\Ft$.  A
\emph{homogeneous linear equation} for an unknown vector $x\in\Ft^\beta$ is
an equation $x\cdot y=0$ with a given vector $y$.  The set $\Ft$ with its two
operations is a field.  Hence the rules of linear algebra hold for vectors
over $\Ft$, and in particular the following fact holds.

\begin{lemma}[Linear equations over $\Ft$]
\label{lem:linear}
A system of fewer than $\beta$ homogeneous linear equations for an unknown
vector $x\in\Ft^\beta$ has a solution $x$ that is different from the zero
vector.
\end{lemma}
\begin{proof}
The solutions of the system form a linear subspace of $\Ft^\beta$.  Its
dimension is at least $\beta$ minus the number of equations, which is at
least $1$.  Hence the subspace contains a vector other than the zero vector.
\end{proof}

The cyclic port labelling gives no information to an agent.  In order to
state this fact, fix an algorithm, a label and an input, and let $q\ge2$.
Consider an agent that observes the degree $q-1$ in every round and that,
after every action $p\ge1$, enters by port $q-p$.  The action of the agent
in a round is a function of its label, its input and its observations in
the previous rounds.  These observations are determined by the previous
actions.  Hence the actions of such an agent form a fixed infinite sequence
of integers in $\{0,\dots,q-1\}$.  We call it the \emph{solo sequence} of
the agent for $q$.  The solo sequence is defined without any graph.  For a
solo sequence $w$, we again denote by $w[1..t]$ the word formed by its
first $t$ actions.

\begin{lemma}[Agents execute fixed words]
\label{lem:words}
Consider an agent in any graph, and let $t\ge0$.  Suppose that the agent
does not meet the other agent during the first $t$ rounds.  Suppose also
that its starting node has degree $q-1$ and that, in each of these rounds
in which it takes a port $p$, it enters a node of degree $q-1$ by port
$q-p$.  Then the actions of the agent in the rounds $1,\dots,t+1$ are the
first $t+1$ actions of its solo sequence for $q$.
\end{lemma}
\begin{proof}
We prove the lemma by induction on the rounds.  Prior to the meeting, an
agent has no information about the other agent.  Hence the action of the
agent in a round $r\le t+1$ is a function of its label, its input and its
observations in the rounds before $r$.  By the inductive hypothesis, its
actions in these rounds are the first $r-1$ actions of the solo sequence.
By the assumptions, its observations in these rounds are those in the
definition of the solo sequence.  Hence its action in round $r$ is action
$r$ of the solo sequence.
\end{proof}

By \cref{lem:words}, an agent executes a fixed word as long as it observes
what it would observe in $K_q$.  In
$K_q$, the walk of a word $w$ from node $x$ is at node $x+w_1+\dots+w_t$
after $t$ rounds.

\section{The lower bound}
\label{sec:quartic}

\begin{figure}[ht]
\centering
\begin{tikzpicture}[x=1cm,y=1cm,
  nd/.style={circle,draw,fill=white,inner sep=0.6pt,minimum size=8.5mm,font=\scriptsize},
  blk/.style={draw,rounded corners=3pt,fill=black!4}]
\foreach \name/\x in {1/0,2/7.4}{
  \node[blk,minimum width=4.8cm,minimum height=4.2cm] at (\x+2.1,1.0) {};
  \node[font=\small] at (\x+2.1,-1.4) {part $P_{\name}$, from a cover of $K_{q_{\name}}$};
}
\node[font=\scriptsize,anchor=east] at (-0.45,2.3) {sheet 1};
\node[font=\scriptsize,anchor=east] at (-0.45,0.3) {sheet 0};
\node[nd] (as) at (0.7,0.3) {$0,0$};
\node[nd] (au) at (2.5,2.3) {$u_1,1$};
\node[nd] (av) at (4.0,2.3) {$v_1,1$};
\node[nd] (au0) at (2.5,0.3) {$u_1,0$};
\node[nd] (av0) at (4.0,0.3) {$v_1,0$};
\node[nd] (bs) at (11.3,0.3) {$0,0$};
\node[nd] (bu) at (9.5,2.3) {$u_2,1$};
\node[nd] (bv) at (8.0,2.3) {$v_2,1$};
\node[nd] (bu0) at (9.5,0.3) {$u_2,0$};
\node[nd] (bv0) at (8.0,0.3) {$v_2,0$};
\draw (au0)--(av0); \draw (bu0)--(bv0);
\draw[dotted,thick] (au)--(av); \draw[dotted,thick] (bu)--(bv);
\draw[very thick,dashed] (av) to[bend left=18] (bv);
\draw[very thick,dashed] (au) to[bend left=38] (bu);
\node[font=\scriptsize] at (0.9,-0.55) {start of agent 1};
\node[font=\scriptsize] at (11.1,-0.55) {start of agent 2};
\end{tikzpicture}
\caption{The graph on which the agents do not meet.  Each part is a
two-sheet cover of a clique from which one edge is removed.  A node is shown
with its projection and its sheet.  The dotted edges are the removed copies,
in sheet $1$, of the hidden edges $\{u_1,v_1\}$ and $\{u_2,v_2\}$.  The
dashed edges are the two new edges.  Agent $j$ traverses only the copy of
$\{u_j,v_j\}$ in sheet $0$, so it never leaves part $P_j$.}
\label{fig:splice}
\end{figure}
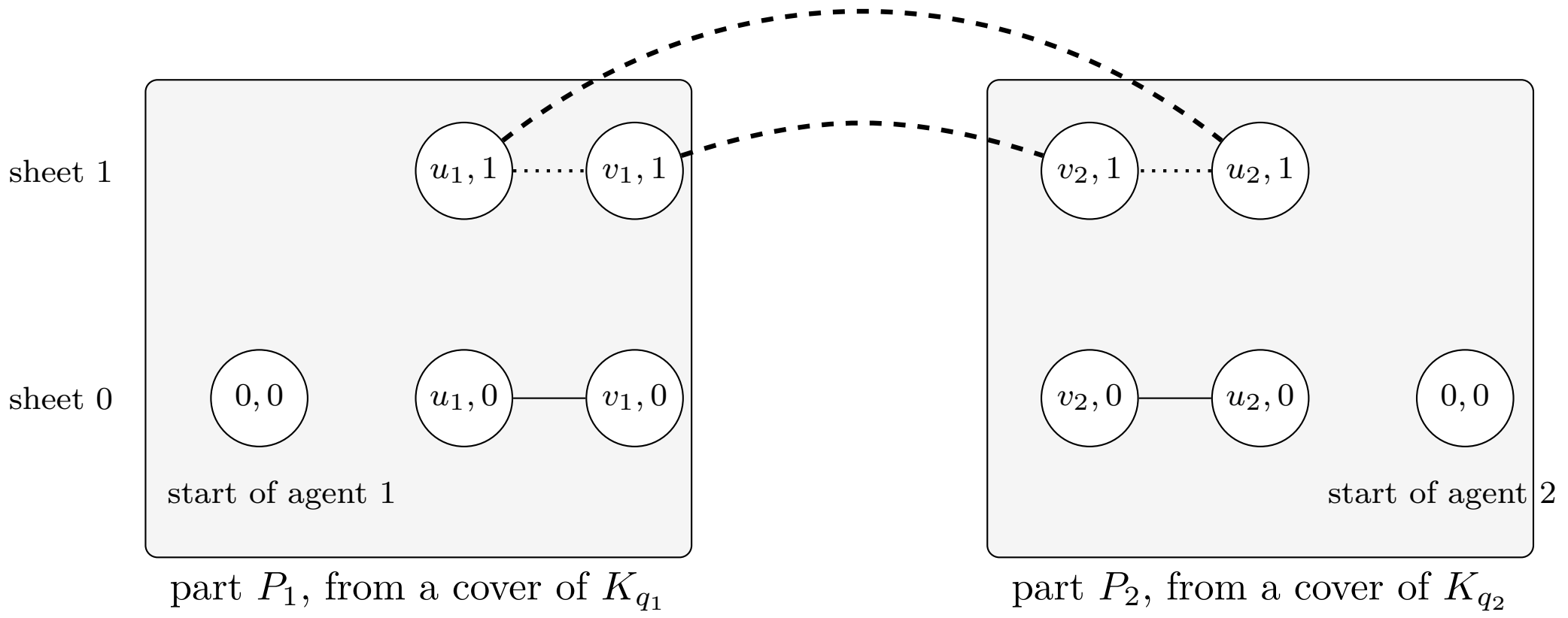

In this section, we prove \cref{thm:quartic}.  The main idea of the proof is the following.  We consider two-sheet covers of the clique $K_q$
with the cyclic port labelling.  By \cref{lem:words}, an agent executes a
fixed word in all such covers.  Suppose that the word has few moves.  Then
some edge $e$ of $K_q$ is traversed by the walk of the word only a few
times.  We choose the signing in such a way that the agent is in sheet $0$
whenever it traverses a copy of $e$.  Hence the agent never traverses the
copy of $e$ in sheet $1$, and we say that this copy is \emph{hidden} from
the agent.  We hide an edge from each agent in a separate cover.  Then we
remove the two hidden copies and join their four ends by two new edges
between the covers (\cref{fig:splice}).  Each agent stays in its own cover
and observes nothing new.  Hence the agents do not meet.

We first show how to hide an edge from one word (\cref{lem:hide}).  Next,
we show that a cover remains connected when the hidden copy is removed
(\cref{lem:bridgeless}).  Then we join two covers and prove the theorem.
We conclude the section by extensions of the lower bound to a family of
covers, to sparse graphs, to one agent, to other models and to teams of
agents (\cref{sec:extensions}), and by the lower bound in terms of the
labels (\cref{sec:labels}).

\subsection{Hiding an edge from one agent}

\begin{lemma}[Hidden edge~\cite{BRT92}]
\label{lem:hide}
Let $q\ge3$ and let $w$ be a word with actions in $\{0,\dots,q-1\}$ and
fewer than $s\beta$ moves.  There
exist an edge $\{u,v\}$ of $K_q$ and a signing $\sg$ with the following
properties.
\begin{enumerate}[label=(\roman*)]
\item The signing $\sg$ is not identically $0$, and $\sg(e)=0$ for every edge
$e$ incident to $u$.
\item The cover $C_\sg$ is connected.
\item The walk of $w$ from $(0,0)$ in $C_\sg$ never traverses the edge
$\{(u,1),(v,1)\}$.
\end{enumerate}
\end{lemma}
\begin{proof}
Consider the walk of $w$ from node $0$ in $K_q$.  It makes fewer than
$s\beta$ traversals, and $K_q$ has $s$ edges.  Hence some edge $\{u,v\}$ is
traversed fewer than $\beta$ times.  The $q-1$ edges incident to $u$ form a spanning
tree of $K_q$ that contains $\{u,v\}$.  We set $\sg(e)=0$ for every edge $e$
of this tree.  The values of $\sg$ on the other $\beta$ edges are unknown
bits, which we regard as a vector $\sg\in\Ft^\beta$.

For every $t\ge0$, let $\chi(t)\in\Ft^\beta$ be the following vector.  Its
coordinate for an edge $e$ outside the tree is the parity of the number of
traversals of $e$ in rounds $1,\dots,t$.  Only the edges outside the tree
can change the sheet of the agent.  Hence, after round $t$, the walk of $w$
from $(0,0)$ in $C_\sg$ is in sheet $\sg\cdot \chi(t)$, where the dot denotes
the scalar product in $\Ft^\beta$.

For every round $t$ in which the walk traverses $\{u,v\}$, we impose the
equation $\sg\cdot \chi(t-1)=0$.  We obtain fewer than $\beta$ homogeneous
linear equations for the unknown vector $\sg\in\Ft^\beta$.  By
\cref{lem:linear}, the system has a solution $\sg$ that is different from
the zero vector, which gives (i).  By the equations, the agent is in sheet $0$
before every traversal of a copy of $\{u,v\}$.  Since $\sg(uv)=0$, it
traverses the copy $\{(u,0),(v,0)\}$.  Hence (iii) holds.  For (ii), the
copies of the tree edges join all nodes of each sheet, and the copies of
an edge $e$ with $\sg(e)=1$ join the two sheets.
\end{proof}

\begin{lemma}[Removing one edge of a cover]
\label{lem:bridgeless}
Let $q\ge3$, let $C_\sg$ be a connected two-sheet cover of $K_q$, and let
$\{x,y\}$ be an edge of $C_\sg$.  The graph obtained from $C_\sg$ by
removing the edge $\{x,y\}$ is connected and has diameter at most $5$.
Moreover, for every node $z$, the distances from $z$ to $x$ and from $z$ to
$y$ in this graph sum to at most $6$.
\end{lemma}
\begin{proof}
Every edge of $K_q$ lies on a triangle.  The walk around a triangle of
$K_q$, started in $C_\sg$, returns to its starting node if the three values
of $\sg$ sum to $0$ in $\Ft$.  Otherwise it returns after going around twice.  Hence
every edge of $C_\sg$ lies on a cycle of length $3$ or $6$.  It follows that
the graph without the edge $\{x,y\}$ is connected, and that it contains a
path of length at most $5$ from $x$ to $y$.

Next, we show that $C_\sg$ has diameter at most $3$.  The graph $C_\sg$ has
$2q$ nodes, and every node has degree $q-1$.  Hence the closed neighbourhood of
a node, which consists of the node and its neighbours, has $q$ nodes.
Consider two nodes that are not adjacent and have no common neighbour.
Their closed neighbourhoods are disjoint and
contain all $2q$ nodes.  Since $C_\sg$ is connected, some edge joins the two
closed neighbourhoods, which gives a path of length $3$ between the two nodes.

From now on, distances are computed in the graph without the edge $\{x,y\}$.
In this graph, the closed neighbourhoods of $x$ and of $y$ have $q-1$
nodes, and the closed neighbourhood of any other node has $q$ nodes.
Consider two nodes at distance $r\ge3$.  Their closed neighbourhoods are
disjoint.  A shortest path between them has $r-1$ inner nodes, and only the
first and the last of these belong to the two closed neighbourhoods.  Hence
at least $r-3$ nodes are outside both closed neighbourhoods.  Since the
graph has $2q$ nodes, we obtain $r\le3$ if neither node is $x$ or $y$, and
$r\le4$ if exactly one of the nodes is $x$ or $y$.  Recall that the distance
between $x$ and $y$ is at most $5$.  Hence the diameter is at most $5$.

Finally, let $z$ be a node.  If $z$ is $x$ or $y$, the two distances sum to
at most $5$.  If both distances are at most $3$, they sum to at most $6$.
Consider the remaining case, in which $z$ is at distance $4$ from one of the
two nodes, say from $x$.  In $C_\sg$, a shortest path from $z$ to $x$ has
length at most $3$.  Hence it uses the edge $\{x,y\}$, and this edge is its
last edge.  The rest of the path joins $z$ to $y$ and has length at most
$2$.  Thus the two distances sum to at most $4+2=6$.
\end{proof}

\subsection{Joining two covers}

We now define the graphs on which the agents do not meet.  Let $q_1=q$ and
$q_2=\lfloor n/2\rfloor-q_1$.  Hence $q_2\in\{q,q+1\}$, and $2q_1+2q_2$ is
equal to $n$ if $n$ is even and to $n-1$ if $n$ is odd.  For $j\in\{1,2\}$
let $L_j=\binom{q_j}2\binom{q_j-1}2$, which is the product $s\beta$ for the
clique $K_{q_j}$.  Thus $L_1=L(n)$ and $L_2\ge L(n)$.

Let $w_1$ be a word with actions in $\{0,\dots,q_1-1\}$ and fewer than
$L_1$ moves.  Let $w_2$ be a word with actions in $\{0,\dots,q_2-1\}$ and
fewer than $L_2$ moves.  The graph $G(w_1,w_2)$ is constructed as follows.
\begin{enumerate}
\item For $j\in\{1,2\}$, apply \cref{lem:hide} to the word $w_j$ in the
clique $K_{q_j}$.  Let $\{u_j,v_j\}$ be the resulting edge and $\sg_j$ the
resulting signing.
\item Let the \emph{part} $P_j$ be the cover $C_{\sg_j}$ without the edge
$\{(u_j,1),(v_j,1)\}$.  The parts $P_1$ and $P_2$ have disjoint sets of
nodes.
\item Add an edge between the node $(u_1,1)$ of $P_1$ and the node
$(u_2,1)$ of $P_2$.  Add an edge between the node $(v_1,1)$ of $P_1$ and
the node $(v_2,1)$ of $P_2$.  At each of its ends, a new edge gets the port
number of the removed edge.
\item If $n$ is odd, replace the first new edge by a path of two edges through a
new node.  The new node has degree $2$, and its ports are numbered
arbitrarily.
\end{enumerate}
The starting node of the agent with label $\ell_j$ is the node $(0,0)$ of
$P_j$.  We call this agent \emph{agent $j$}.

\begin{lemma}
\label{lem:splice}
The graph $G(w_1,w_2)$ is simple and connected, has size $n$ and has
diameter at most $7$.  It is $(q-1)$-regular if $n$ is divisible by
$4$.  Suppose that agent $1$ executes $w_1$ and agent $2$ executes $w_2$ from
their starting nodes.  Then agent $j$ stays in $P_j$ while it executes $w_j$, and the agents do
not meet while both of them execute their words.  Moreover, agent $j$ sees the degree $q_j-1$ at every node, and it
enters every node by port $q_j-p$ after taking port $p$.
\end{lemma}
\begin{proof}
By \cref{lem:bridgeless}, both parts are connected.  The new edges join the
two parts.  Hence the graph is connected.  The graph is simple, because the new
edges join different parts.  Its size is $n$, because the parts have
$2q_1+2q_2$ nodes and one node is added if $n$ is odd.  Every node of a
part keeps its degree and its port numbers.  If $n$ is divisible by $4$,
then $q_1=q_2=q$ and no node is added, so every node has degree $q-1$.

We now bound the diameter.  By \cref{lem:bridgeless}, two nodes of the same
part are at distance at most $5$.  Consider a node $z_1$ of $P_1$ and a
node $z_2$ of $P_2$.  One path from $z_1$ to $z_2$ leads through
$(u_1,1)$ and $(u_2,1)$, and another path leads through $(v_1,1)$ and
$(v_2,1)$.  By \cref{lem:bridgeless}, the edges of the two paths inside
$P_j$ number at most $6$, for each $j$.  The two paths contain two or three
further edges.  Hence their lengths sum to at most $15$, and one of them
has length at most $7$.  If $n$ is odd, the new node is adjacent to
$(u_1,1)$ and to $(u_2,1)$.  Hence it is at distance at most $1+5$ from
every other node.

By \cref{lem:hide}(iii), the walk of $w_j$ in $C_{\sg_j}$ never traverses
the removed edge.  Hence it is also a walk in $P_j$, and it takes no port of
a new edge.  At every node the agent sees the degree $q_j-1$, and it enters
by port $q_j-p$ after taking port $p$.  The two agents are always in
different parts, so they do not meet.
\end{proof}

\subsection{Proof of the lower bound}

\begin{proof}[Proof of \cref{thm:quartic}]
Fix an algorithm and two distinct labels $\ell_1$ and $\ell_2$.  Let both
agents receive the same input, which may contain $n$ and both labels.  For
$j\in\{1,2\}$, let $w_j$ be the solo sequence of agent $j$ for $q_j$.  The
sequences $w_1$ and $w_2$ are determined by the algorithm, the labels and
the input.  We say that an integer $t\ge0$ is \emph{admissible} if $w_1[1..t]$
has fewer than $L_1$ moves and $w_2[1..t]$ has fewer than $L_2$ moves.  For
an admissible $t$, let $G_t$ denote the graph $G(w_1[1..t],w_2[1..t])$.

\emph{The agents follow their solo sequences.}  Let $t$ be admissible.  We claim
that, in the graph $G_t$, the agents do not meet during the rounds
$1,\dots,t$, and agent $j$ makes the first $t+1$ actions of $w_j$ in the
rounds $1,\dots,t+1$.  Indeed, let $r\le t$ and suppose that each agent $j$
has made the first $r$ actions of $w_j$ in the rounds $1,\dots,r$.  By
\cref{lem:splice}, the two agents have been in different parts after each
of these rounds, each agent $j$ has seen the degree $q_j-1$ at every node,
and it has entered every node by port $q_j-p$ after taking port $p$.  By
\cref{lem:words}, agent $j$ makes action $r+1$ of $w_j$ in round $r+1$.
The claim follows by induction on $r$.

\emph{Some integer is not admissible.}  Suppose, for a contradiction, that every
$t\ge0$ is admissible.  By the claim, for every $t$ there is a graph of size $n$
with two starting nodes in which the agents do not meet during the first
$t$ rounds.  There are finitely many port-labelled graphs of size $n$.
Hence one graph, with the same two starting nodes, serves for infinitely
many values of $t$.  In this graph the agents never meet, which contradicts
the correctness of the algorithm.

\emph{The bound.}  The integer $0$ is admissible, and if $t$ is admissible
then so is every smaller integer.  By the previous step, there is a largest
admissible integer, which we denote by $T$.  Since $T+1$ is not admissible,
for some $j\in\{1,2\}$ the word $w_j[1..T+1]$ has at least $L_j$ moves.
Since $T$ is admissible, the word $w_j[1..T]$ has fewer than $L_j$ moves.
Hence action $T+1$ of $w_j$ is a move, and it is move number $L_j$ of $w_j$.

Consider the graph $G_T$.  By the claim, the agents do not meet during the
rounds $1,\dots,T$, and agent $j$ makes the first $T+1$ actions of $w_j$ in
the rounds $1,\dots,T+1$.  Thus the agents do not meet before the end of
round $T+1$, and agent $j$ traverses $L_j$ edges during the rounds
$1,\dots,T+1$.  Hence agent $j$ traverses at least $L_j\ge L(n)$ edges until
the meeting, which proves (b).  An agent traverses at most one edge per
round.  Thus $T+1\ge L_j\ge L(n)$, and rendezvous takes at least $L(n)$
rounds, which proves (a).
By \cref{lem:splice}, the graph $G_T$ has diameter at most
$7$, and it is $(q-1)$-regular if $n$ is divisible by $4$.

The two solo sequences are defined separately.  Hence the proof remains
valid if the two agents execute different algorithms.
\end{proof}

\subsection{Extensions}
\label{sec:extensions}

The first extension shows that the lower bound $\Omega(n^4)$ holds inside
the family of
\cref{cor:fourcover}.  It costs a factor $2$ in the bound.

\begin{corollary}[Lower bound on four-sheet covers of the clique]
\label{cor:quartic-fourcover}
Let $q\ge3$ and $n=4q$.  For any algorithm and any two distinct labels,
there exists a connected four-sheet cover of $K_q$ in which agents that
start simultaneously do not meet during the first
$\lceil s\beta/2\rceil-1$ rounds.  The statement holds even if the agents
know $q$.
\end{corollary}
\begin{proof}
Let $T=\lceil s\beta/2\rceil-1$, so that $2T<s\beta$.  For $j\in\{1,2\}$,
let $w_j$ be the word formed by the first $T$ actions of the solo sequence
of agent $j$ for $q$.  The walks of the two words from node $0$ in $K_q$
together traverse some edge $\{u,v\}$ of $K_q$ fewer than
$\beta$ times.  Hence each of the two words traverses it fewer than $\beta$
times.  We repeat the proof of \cref{lem:hide} for each word with this
common edge $\{u,v\}$, and obtain signings $\sg_1$ and $\sg_2$.  We define
the parts as before.  We add an edge between $(u,1)$ of $P_1$ and $(v,1)$
of $P_2$, and an edge between $(v,1)$ of $P_1$ and $(u,1)$ of $P_2$.  Every
edge of the resulting graph has the port numbers of an edge of $K_q$
between the projections of its ends.  Hence the graph is a four-sheet cover
of $K_q$.  It is connected by \cref{lem:bridgeless}.  As in
\cref{lem:splice}, an agent $j$ that executes $w_j$ from the node $(0,0)$
of $P_j$ stays in $P_j$.  By \cref{lem:words} and induction on the rounds,
as in the proof of \cref{thm:quartic}, agent $j$ executes $w_j$ in the
first $T$ rounds.  Hence the agents do not meet during these rounds.
\end{proof}

The second extension concerns sparse graphs.  A \emph{bridge} is an edge
whose removal disconnects the graph, and a graph without a bridge is called
\emph{bridgeless}.  A \emph{circulant graph} has the nodes $0,\dots,q-1$
and a given set of offsets.  It has an edge between $v$ and $v+x$ modulo
$q$, for every node $v$ and every offset $x$ of the set.

\begin{remark}[Sparse graphs]
\label{rem:edges}
\Cref{lem:hide} and the connectivity in \cref{lem:bridgeless} use only four
properties of the base.  It is connected, bridgeless and regular, and
the port by which an agent enters a node is a function of the port that it
took.  For such a base, the star of the edges incident to $u$ is extended to a
spanning tree, on which the signing is $0$.  The number of unknown bits
is the cycle rank $\beta$ of the base.  As an example, let
$3\le d\le q-1$, with $q$ even if $d$ is odd.  Take the circulant graph with $q$
nodes and a set of $d$ offsets that contains $1$ and $-1$ and contains
$-x$ for each of its elements $x$.  Let the port numbers correspond to the
offsets.  This base has $s=qd/2$ edges and $\beta=s-q+1>s/3$.  The
construction gives a $d$-regular graph of size $n=4q$, with
$m=4s$ edges.  On this graph rendezvous takes at least $s\beta>m^2/48$
rounds, and one agent traverses the same number of edges.  The bound has the order
$d^2n^2$ of the bound of Borodin et al.~\cite{BRT92} for universal
traversal sequences.
\end{remark}

The third extension concerns one agent.  Recall that in exploration an
agent has to visit all nodes, and that in treasure hunt it has to visit a
node that it does not know, called the treasure.  An \emph{exploration
algorithm} is an algorithm with which the agent visits all nodes, in every
graph and from every starting node.  The following corollary is
the argument of Borodin et al.~\cite{BRT92} in our model.

\begin{corollary}[One agent]
\label{cor:treasure}
Let $n\ge7$ and $q=\lfloor(n-1)/2\rfloor$.  For any deterministic
exploration algorithm of one agent, there exist a graph of size $n$, a
starting node and a node $z$ such that the agent traverses at least
\[
 \binom q2\binom{q-1}2=(1-o(1))\,\frac{n^4}{64}
\]
edges until it visits $z$.  The statement holds even if the agent knows
$n$.  Hence exploration takes at least the same number of rounds, and so does
treasure hunt in the worst case over the positions of the treasure.
\end{corollary}
\begin{proof}
Let $w$ be a word with actions in $\{0,\dots,q-1\}$ and fewer than
$s\beta$ moves.  Apply \cref{lem:hide} to
$w$, and replace the edge $\{(u,1),(v,1)\}$ of $C_\sg$ by a path through
$n-2q$ new nodes.  There are one or two new nodes, and their ports are
numbered arbitrarily.  At the nodes $(u,1)$ and $(v,1)$, the path gets the
port numbers of the removed edge.  Let $z$ be a new node.  The resulting graph is
connected and has size $n$.  By \cref{lem:hide}(iii), an agent that executes
$w$ from $(0,0)$ never visits a new node.  Hence it sees the degree $q-1$
at every node and enters by port $q-p$ after taking port $p$.

Let $w$ now denote the solo sequence of the agent for $q$.  As in the proof
of \cref{thm:quartic}, we consider the integers $t$ for which $w[1..t]$ has
fewer than $s\beta$ moves, and the graph constructed for the word $w[1..t]$.
By \cref{lem:words}, in this graph the agent does not visit $z$ during the
first $t$ rounds, and it makes action $t+1$ of $w$ in round $t+1$.  The
number of graphs of size $n$ is finite, and the agent has to visit $z$.
Hence there is a largest such integer $t$.  In the graph constructed for
it, the agent makes its move number $s\beta$ in round $t+1$, and it has not
visited $z$ before this round.
\end{proof}

The fourth extension concerns other models.

\begin{remark}[Gathering and asynchronous agents]
\label{rem:models}
\Cref{thm:quartic} uses only two properties of the agents.  Before the
meeting, the actions of an agent depend only on its label, its input and
its observations.  The agents have to meet whenever they start
simultaneously at two nodes of a connected graph of size $n$.  Both
properties hold for an algorithm that gathers every team of at least two
agents, when the team consists of two agents.  Hence such an algorithm
needs at least $L(n)$ rounds on some graph of size $n$.

Recall that in the asynchronous model an adversary chooses the speed of each
agent, and the cost is the total number of edge traversals.  In this model
the agents may also meet inside an edge.  Let the adversary start the agents simultaneously and
let each agent traverse one edge per round.  In the graph $G(w_1,w_2)$ the
two agents are then in different parts, so they do not meet inside an edge
either.  We repeat the proof of \cref{thm:quartic} with this adversary.  For an
admissible $t$, the agents stay in different parts of $G_t$ during the first
$t$ rounds, so they meet neither at a node nor inside an edge.  The
algorithm guarantees a meeting, so there is a largest admissible integer
$T$, and in round $T+1$ one agent makes its move number $L_1$ or $L_2$ in
$G_T$.  Suppose that the agents meet inside an edge in this round.  Then
the edge is a new edge, and the other agent has made at least one move,
because it is at a node of sheet $1$.  Hence the agents traverse at least
$L(n)$ edges in total until the meeting, and every asynchronous rendezvous
algorithm has cost at least $L(n)$ on some graph of size $n$.
\end{remark}

The fifth extension concerns teams of more than two agents.  We consider
$k$ agents with distinct labels that have to gather at one node.  Before an
agent meets another agent, its actions depend only on its label, its input
and its observations.

\begin{corollary}[Teams of $k$ agents]
\label{cor:teams}
Let $k\ge2$, let $n\ge6k$ and let $q=\lfloor n/(2k)\rfloor$.  For any
deterministic gathering algorithm for $k$ agents and any $k$ distinct
labels, there exist a graph of size $n$ and $k$ starting nodes such that,
if the agents start simultaneously, then no two agents meet during the
first $\binom q2\binom{q-1}2-1$ rounds.  The statement holds even if the
agents know $n$, $k$ and all labels.  Hence gathering $k$ agents takes
$\Omega(n^4/k^4)$ rounds.
\end{corollary}
\begin{proof}
Let $w_1,\dots,w_k$ be words with actions in $\{0,\dots,q-1\}$ and fewer
than $s\beta$ moves each.  For every $j\in\{1,\dots,k\}$, apply
\cref{lem:hide} to $w_j$, and let $P_j$ be the cover $C_{\sg_j}$ without
the edge $\{(u_j,1),(v_j,1)\}$.  For every $j$, add an edge between the
node $(v_j,1)$ of $P_j$ and the node $(u_{j+1},1)$ of $P_{j+1}$, where
$P_{k+1}$ denotes $P_1$.  At each of its ends, a new edge gets the port
number of the removed edge.  If $2qk<n$, replace one new edge by a path
through $n-2qk$ new nodes, whose ports are numbered arbitrarily.  The
resulting graph is simple and has size $n$.  It is connected, because the
parts are connected by \cref{lem:bridgeless}.  Agent $j$ starts at the node
$(0,0)$ of $P_j$.  As in \cref{lem:splice}, an agent $j$ that executes
$w_j$ stays in $P_j$, sees the degree $q-1$ at every node, and enters by
port $q-p$ after taking port $p$.  Hence no two agents meet.

We now take the solo sequences of the $k$ agents for $q$, and we argue as
in the proof of \cref{thm:quartic}.  We consider the integers $t$ for which
the first $t$ actions of each solo sequence contain fewer than $s\beta$
moves, and the graph constructed for these $k$ words.  In this graph, no
two agents meet during the first $t$ rounds, and every agent makes action
$t+1$ of its solo sequence in round $t+1$.  If every $t$ had this property,
then no two agents would ever meet in some graph of size $n$, and the
agents would not gather.  Hence there is a largest such integer $T$.  In
the graph constructed for it, some agent makes its move number $s\beta$ in
round $T+1$.  Thus $T+1\ge s\beta$.
\end{proof}

For $k=2$ the bound of \cref{cor:teams} is the bound of \cref{thm:quartic}.
The bound decreases with $k$, because each agent is enclosed in a separate
part of about $n/k$ nodes.  We do not know whether the factor $1/k^4$ is
necessary.

We conclude the extensions with a remark on the constant of the lower
bound.

\begin{remark}[The hidden edge lemma is tight]
\label{rem:hide-tight}
\Cref{lem:hide} hides an edge from every word with fewer than $s\beta$
moves.  This number cannot be raised by more than the factor $1+O(1/q)$.
Indeed, there is a word of length at most $(\beta+1)(s+q/2)+3\beta$ whose
walk traverses every edge, in every connected two-sheet cover of $K_q$ and
from every node.  We sketch its construction.  Let $w$ be a word whose walk
from node $0$ in $K_q$ traverses every edge and returns to node $0$.  For
odd $q$ all degrees of $K_q$ are even, and we take a closed walk that
traverses every edge once.  For even $q$ we take a closed walk that
traverses the edges of one perfect matching twice and every other edge
once.  Hence $w$ has at most $s+q/2$ actions.  Let the spanning tree
consist of the edges incident to node $0$, and let $\chi\in\Ft^\beta$ be
the vector of the parities of the traversals of the other $\beta$ edges by
$w$.  For an edge $\{u,v\}$ with $0<u<v$, the three actions $u$, $v-u$ and
$q-v$ lead from node $0$ through $u$ and $v$ back to node $0$.  We call
them the triangle of the edge.  The word $w$ followed by the triangle of an
edge has the vector $\chi$ with the coordinate of this edge changed.  These
$\beta$ vectors and the vector $\chi$ together contain $\beta$ vectors such
that every vector of $\Ft^\beta$ is a sum of some of them.  The long word
consists of $\beta+1$ copies of $w$.  Between consecutive copies we insert
a triangle or nothing, so that the $\beta$ chosen vectors occur.  Consider
a connected two-sheet cover.  Exchanging the names of the two copies of
some nodes, we may assume that its signing is $0$ on the edges incident to
node $0$.  The signing is not identically $0$, so its scalar product with
one of the chosen vectors is $1$.  Hence two consecutive copies of $w$
start at the two copies of node $0$.  Their walks are in different sheets
after every action, so together they traverse both copies of every edge.
By the cyclic port labelling, the same holds from every starting node.
Thus a better constant in \cref{thm:quartic} requires an argument that does
more than hide an edge of a two-sheet cover of $K_q$.  Borodin et
al.~\cite{BRT92} make the corresponding observation for their bound.  They
note that, for the covers of a fixed graph with a fixed starting node and a
fixed spanning tree, a sequence of length $O(d^2n^2)$ traverses all edges.
\end{remark}

\subsection{The length of the labels}
\label{sec:labels}

The bound of \cref{thm:quartic} does not depend on the labels.  We now prove
\cref{thm:labels}, which shows that the time of rendezvous grows with the
length of the labels.  The main idea of the proof is the following.  We
use two-sheet covers of $K_q$ in which exactly one edge $f$ has value $1$.
The agents start at the two copies of node $0$.  Thus they start in
different sheets, and an agent changes its sheet only when it traverses a
copy of $f$.  Suppose that the two agents traverse copies of $f$ in exactly
the same rounds.  Then they are in different sheets after every round, and
they never meet.  A short word traverses most edges only a few times.  There
are few ways to choose a few rounds.  Hence, among many labels, two labels
traverse some edge $f$ in the same rounds.

\begin{proof}[Proof of \cref{thm:labels}]
Fix an algorithm and let all agents receive the same input, which may
contain $n$ and $\Lambda$.  For an edge $f$ of $K_q$, let $C_f$ denote the
two-sheet cover of $K_q$ whose signing has value $1$ on $f$ and value $0$ on
all other edges.  The graph $C_f$ has size $2q=n$.  It is connected, because
each sheet is a copy of $K_q$ without one edge, and the two copies of $f$
join the sheets.  In $C_f$ we let the agents start simultaneously at the
nodes $(0,0)$ and $(0,1)$.

Let $T=\lceil(k+1)s/2\rceil-1$.  For a label $\ell$, let $w_\ell$ be the
word formed by the first $T$ actions of the solo sequence for $q$ of the
agent with label $\ell$.  Every node of every graph $C_f$ has degree $q-1$,
and every edge has the port numbers of an edge of $K_q$.  Hence, by
\cref{lem:words}, the agent with label $\ell$ executes $w_\ell$ in every
graph $C_f$, as long as it does not meet the other agent.  For an edge $f$
of $K_q$, let
$\tau_\ell(f)$ be the set of rounds in which the walk of $w_\ell$ from node
$0$ in $K_q$ traverses $f$.

We first show that agents with labels $\ell$ and $\ell'$ do not meet in
$C_f$ during the first $T$ rounds if $\tau_\ell(f)=\tau_{\ell'}(f)$.
Indeed, an agent changes its sheet exactly in the rounds in which it
traverses a copy of $f$.  These rounds are the same for both agents, and
the agents start in different sheets.  Hence they are in different sheets
after every round.

Next, we find two such labels.  The word $w_\ell$ traverses at most
$T<(k+1)s/2$ edges.  Hence fewer than $s/2$ edges of $K_q$ are traversed
more than $k$ times by its walk from node $0$.  We call the other edges
\emph{light} for $\ell$.  Every label has more than $s/2$ light edges, and
$K_q$ has $s$ edges.  Thus some edge $f$ is light for more than $L/2$ labels
of $\Lambda$.  For each of these labels, $\tau_\ell(f)$ is a set of at most
$k$ rounds among the rounds $1,\dots,T$.  If $k=0$, there is one such set.
If $k\ge1$, the number of such sets is at most
$(\mathrm eT/k)^k\le(\mathrm es)^k$, because $T/k<(k+1)s/(2k)\le s$.  By
the choice of $k$ we have $(\mathrm es)^k\le L/2$.  Hence two of the labels
have the same set $\tau_\ell(f)$, and agents with these labels do not meet
in $C_f$ during the first $T$ rounds.

It remains to compute the distances in $C_f$.  By the proof of
\cref{lem:bridgeless}, the diameter of $C_f$ is at most $3$.  The nodes
$(0,0)$ and $(0,1)$ are not adjacent.  A path of length $2$ between them
would traverse the two copies of one edge incident to node $0$, which have
no common end.  Hence the initial distance is at least $3$, and both the
initial distance and the diameter are equal to $3$.
\end{proof}

The set of all labels of length $b$ has $2^{b-1}$ elements.  For this set,
$k+1>(b-2)/\log(\mathrm es)$ and $\log(\mathrm es)<2\log n$.  Hence
\cref{thm:labels} gives the following bound.

\begin{corollary}
\label{cor:labels}
Let $n\ge6$ be even and let $b\ge3$.  For any deterministic rendezvous
algorithm, there exist two labels of length $b$, a graph of size $n$ and
two starting nodes such that agents that start simultaneously do not meet
during the first $T$ rounds, for an integer
$T\ge\frac{n(n-2)(b-2)}{32\log n}-1$.
\end{corollary}

Together with \cref{thm:quartic}, for even $n\ge12$ the time of rendezvous
for labels of length $b$ is $\Omega(n^4+n^2b/\log n)$ in the worst case.

\begin{remark}[Odd sizes divisible by three]
\label{rem:three}
The proof of \cref{thm:labels} extends to every $n\ge9$ that is divisible
by $3$.  Let $q=n/3$.  For an edge $f=\{u,v\}$ of $K_q$ with $u<v$, we take
the three-sheet cover of $K_q$ in which the copies of $f$ join the node
$(u,a)$ to the node $(v,a+1)$, with the sheets taken modulo $3$, and the
copies of every other edge stay inside the sheets.  This cover is connected
and has size $n$.  The agents start at $(0,0)$ and $(0,1)$.  An agent that
traverses a copy of $f$ from $u$ to $v$ moves to the next sheet, and an
agent that traverses it from $v$ to $u$ moves to the previous sheet.  For a
label we now record the rounds in which its word traverses $f$, together
with the direction of each traversal.  Two agents with the same record are
in different sheets after every round.  The number of records with at most
$k\ge1$ rounds is at most $(2\mathrm eT/k)^k$.  Hence, for
$k=\lfloor(\log L-1)/\log(2\mathrm es)\rfloor$, some two labels of
$\Lambda$ do not meet during the first $\lceil(k+1)s/2\rceil-1$ rounds.
It follows that the bound $\Omega(n^2b/\log n)$ holds for every $n$ that is
divisible by $2$ or by $3$.
\end{remark}

\begin{remark}[The limit of the argument]
\label{rem:labels-limit}
The proof of \cref{thm:labels} uses only the graphs $C_f$ with the starting
nodes $(0,0)$ and $(0,1)$.  These graphs cannot give more than the order
$n^2(1+b/\log n)$.  Indeed, suppose that the agents know $q\ge4$ and know
that they start simultaneously at the two copies of node $0$ in such a
graph.  Recall from \cref{rem:hide-tight} the triangle of an edge $\{u,v\}$ with
$0<u<v$, which consists of the three actions $u$, $v-u$ and $q-v$.  There are $\beta=\binom{q-1}2$ triangles,
and we fix their order.  Every edge of $K_q$ belongs to some triangle, and
no triangle traverses an edge twice.  An agent writes its label in base
$\beta-1$, with the digits $0,\dots,\beta-2$, and appends the digit
$\beta-1$ infinitely often.  For every digit $x$ of the resulting sequence, the
agent stays for $3x$ rounds, then executes all $\beta$ triangles in the
fixed order, and then stays for $3(\beta-x)$ rounds.  Thus every digit
takes $6\beta$ rounds, and after every third round both agents are at a
copy of node $0$.

Consider the first digit in which the sequences of the two agents differ.
It is not later than the first appended digit of the shorter sequence.  For
the earlier digits, the agents traverse copies of $f$ in the same rounds,
so they are in different sheets.  For this digit, let $x<x'$ be the digits
of the two agents.  Consider the first triangle in the fixed order that
contains $f$.  The agent with the digit $x$ completes it earlier than the
other agent.  At this time the agent has changed its sheet once, and the
other agent has not yet traversed a copy of $f$.  Both agents are at a copy
of node $0$, and hence at the same node.  The agents thus meet within
$6\beta(2+b/\log(\beta-1))$ rounds, which is $O(n^2(1+b/\log n))$.  It
follows that a lower bound of order $n^2b$ needs other graphs.
\end{remark}

\section{The lower bound at distance seven}
\label{sec:seven}

The graphs of \cref{thm:quartic} have diameter at most seven, but their
initial distance is not controlled.  In this section, we prove a lower bound of the
same order on graphs in which the diameter and the initial distance are
both equal to seven.  To this end, we keep the signing equal to $0$ on more
edges.  We first prove the corresponding version of \cref{lem:hide}.  Then
we prove the theorem.  We conclude the section by a remark, which explains
why the argument does not work for an initial distance below seven.

\begin{lemma}[Hidden edge far from the starting node]
\label{lem:hide-seven}
Let $q\ge5$ and $\beta'=\binom{q-3}2$, and let $w$ be a word with actions
in $\{0,\dots,q-1\}$ and fewer than
$\binom{q-1}2\beta'$ moves.  There exist an edge $\{u,v\}$ of $K_q$ with
$0\notin\{u,v\}$ and a signing $\sg$ that is not identically $0$, with the
following properties.  The signing $\sg$ is $0$ on every edge incident to
$0$, $u$ or $v$.  The walk of $w$ from $(0,0)$ in $C_\sg$ never traverses
the edge $\{(u,1),(v,1)\}$.  Moreover, the graph obtained from $C_\sg$
by removing this edge satisfies the following conditions.
\begin{enumerate}[label=(\roman*)]
\item Every node is at distance at most $3$ from $(u,1)$ and from
$(v,1)$.
\item The node $(0,0)$ is at distance exactly $3$ from $(u,1)$ and from
$(v,1)$.
\end{enumerate}
\end{lemma}
\begin{proof}
There are $\binom{q-1}2$ edges of $K_q$ that are not incident to $0$.  The walk of $w$
from $0$ traverses them fewer than $\binom{q-1}2\beta'$ times in total.  Hence one of them,
say $\{u,v\}$, is traversed fewer than $\beta'$ times.  Exactly $\beta'$ edges of
$K_q$ have both ends outside $\{0,u,v\}$.  We set $\sg=0$ on all other
edges.  As in the proof of \cref{lem:hide}, we obtain fewer than $\beta'$
homogeneous linear equations for a vector of $\beta'$ unknown bits.  By
\cref{lem:linear}, a solution $\sg$ different from the zero vector exists,
and the walk never traverses $\{(u,1),(v,1)\}$.

We prove (i) for $(u,1)$.  The proof for $(v,1)$ is the same.  Recall that
all edges incident to $0$, $u$ and $v$ have the value $0$.  Consider first
the nodes of sheet $1$.  The nodes $(v,1),(0,1),(u,1)$ form a path, and
$(x,1)$ is adjacent to $(u,1)$ for every $x\notin\{u,v\}$.  Consider now a
node $(x,0)$.  If some edge $\{x,z\}$ has $\sg(xz)=1$, then
$z\notin\{0,u,v\}$ and $(x,0),(z,1),(u,1)$ is a path.  Otherwise we take
any edge $\{y,z\}$ with $\sg(yz)=1$.  Then $x\notin\{y,z\}$, the edge
$\{x,y\}$ has the value $0$, and $(x,0),(y,0),(z,1),(u,1)$ is a path.  None
of these paths uses the removed edge.

We prove (ii).  A path of length $1$ or $2$ from $(0,0)$ to $(u,1)$ or to
$(v,1)$ would consist of copies of edges incident to $0$, $u$ or $v$.
Such copies do not change the sheet.  Hence the distance is at least $3$,
and by (i) it is exactly $3$.
\end{proof}

\begin{theorem}[Lower bound at distance seven]
\label{thm:seven}
Let $n\ge20$ and let $q=\lfloor n/4\rfloor$.  For any deterministic
rendezvous algorithm and any two distinct labels, there exist a graph of
size $n$ and two starting nodes with the following properties.  The
diameter of the graph and the initial distance are both equal to $7$.  If
the agents start simultaneously, then they do not meet during the first
\[
 \binom{q-1}2\binom{q-3}2-1=\frac{(q-1)(q-2)(q-3)(q-4)}4-1
\]
rounds, and one of the agents traverses at least
$\binom{q-1}2\binom{q-3}2$ edges until the meeting.  The statement holds
even if both agents know $n$, both labels, the diameter and the initial
distance.
\end{theorem}
\begin{proof}
We repeat the construction of \cref{sec:quartic}, with
\cref{lem:hide-seven} instead of \cref{lem:hide}.  The numbers $L_1$ and
$L_2$ are replaced by the products $\binom{q_j-1}2\binom{q_j-3}2$ for $j=1$ and
$j=2$, and the smaller of them is the product for $q$.
The proof of \cref{thm:quartic} applies without other changes.  It remains
to compute the distances in $G(w_1,w_2)$.

By \cref{lem:hide-seven}(i), two nodes of the same part $P_j$ are joined
through $(u_j,1)$ by a path of length at most $6$.  Two nodes of different
parts are joined by a path of length at most $3+1+3=7$ that uses the second
new edge.  If $n$ is odd, the new node is adjacent to $(u_1,1)$ and to
$(u_2,1)$, so it is at distance at most $4$ from every other node.  Hence
the diameter is at most $7$.  A path between the two
starting nodes leaves $P_1$ through $(u_1,1)$ or $(v_1,1)$, and it enters
$P_2$ for the last time through $(u_2,1)$ or $(v_2,1)$.  By \cref{lem:hide-seven}(ii), such a
path has at least $3$ edges in $P_1$ before it leaves $P_1$ for the first
time, and at least
$3$ edges in $P_2$ after it enters $P_2$ for the last time.  Hence the initial distance is
at least $7$.  It follows that the initial distance and the diameter are
both equal to $7$.
\end{proof}

\begin{remark}[Why seven]
\label{rem:why-seven}
The proofs of \cref{thm:quartic,thm:seven} separate the agents by edges
that neither agent has traversed.  At an initial distance below seven, such
an argument cannot prove more than $O(n^3)$.  Indeed, an agent may begin by
trying every sequence of three ports and returning after each sequence.
The agent needs at most $6n^3$ rounds for these attempts.  Afterwards it
has traversed every edge that has an end at distance at most $2$ from its
starting node.  Consider an edge that the agent has not traversed after
$6n^3$ rounds.  Both ends of this edge are at distance at least $3$ from
the starting node of the agent.  Suppose that the two agents are still
separated by edges that neither of them has traversed.  Then their initial
distance is at least $3+1+3=7$.
\end{remark}

\section{The upper bounds}
\label{sec:upper}

In this section, we prove \cref{thm:upper} and \cref{cor:fourcover}.  The main idea of the algorithms is the following.  Suppose that one agent stays at its starting node and the other agent
makes a random walk, which means that it chooses its moves at random.  Then the expected time until the walking agent finds the other agent is
bounded by a property of the walk, called its hitting time.  We use
a random walk whose hitting time is $O(n^2)$ in every graph.  A
deterministic agent cannot make random choices.  However, a sufficiently
long sequence of random choices succeeds with large probability in one
graph.  If the graph belongs to a known finite family, then some fixed
sequence succeeds in all graphs of the family simultaneously.  The agents
find such a sequence by exhaustive search and use it instead of the random
choices.  This argument is due to Aleliunas et al.~\cite{AKLLR79}.  The
labels decide how an agent walks and when it stays.

We first describe the random walk and a procedure by which an agent
simulates one step of it (\cref{sec:walk}).  Next, we present two
algorithms for agents that know a finite family of graphs to which the
graph belongs.  The first algorithm is for a simultaneous start
(\cref{sec:sim}), and the second algorithm is for an arbitrary delay
(\cref{sec:delay}).  They prove \cref{cor:fourcover}, and they introduce
the tools of the main algorithm.  We conclude the section with the algorithm for agents
that know nothing (\cref{sec:unknown}), which proves \cref{thm:upper}.

\subsection{A random walk that an agent can simulate}
\label{sec:walk}

A \emph{random walk} on a graph is a random sequence $X_0,X_1,\dots$ of
nodes in which $X_{t+1}$ depends only on $X_t$.  It is given by its
\emph{transition probabilities} $\rho(u,v)$.  The number $\rho(u,v)$ is the
probability that $X_{t+1}=v$ when $X_t=u$.  We use the following walk,
which we call the \emph{lazy Metropolis walk}.  It is the walk given by the
rule of Metropolis et al.~\cite{MRRTT53} for the uniform distribution on the
nodes, with the probability of every move halved.  Nonaka et
al.~\cite{NOSY10} call the walks given by this rule Metropolis walks.  Let $d(u)$ denote the degree
of a node $u$.  For every edge $\{u,v\}$ and every node $u$,
\begin{equation}
 \rho(u,v)=\frac1{2\max\{d(u),d(v)\}},\qquad
 \rho(u,u)=1-\sum_{v\text{ adjacent to }u}\rho(u,v).
 \label{eq:metropolis}
\end{equation}
All other transition probabilities are $0$.  The walk is \emph{lazy}, which
means that $\rho(u,u)\ge\frac12$ for every node $u$.  The walk is
\emph{symmetric}, which means that $\rho(u,v)=\rho(v,u)$ for all nodes $u$ and
$v$.

The \emph{hitting time} $H(x,y)$ is the expected number of steps that the
walk started at node $x$ needs to reach node $y$.  Nonaka et
al.~\cite{NOSY10} proved that the hitting times of Metropolis walks whose
target distribution is uniform are $O(n^2)$.  We include a short proof for the walk \eqref{eq:metropolis}, in
order to keep the paper self-contained.

\begin{lemma}[Hitting time]
\label{lem:hit}
In any connected graph of size $n\ge2$, the hitting times of the lazy
Metropolis walk satisfy $H(x,y)<12n^2$ for all nodes $x$ and $y$.
\end{lemma}
\begin{proof}
We use the correspondence between symmetric random walks and electrical
networks~\cite{CRRST96}.  Regard every edge $\{u,v\}$ as a resistor with
resistance $1/\rho(u,v)$.  The \emph{effective resistance} $R(x,y)$ between the nodes $x$ and $y$
is the voltage between $x$ and $y$ when a current of one unit enters this
network at $x$ and leaves it at $y$.  For a symmetric walk on a graph of size $n$,
the hitting times satisfy the identity $H(x,y)+H(y,x)=n\,R(x,y)$.  The
effective resistance between two nodes is at most the sum of the
resistances along any path that joins them.

Consider a shortest path from $x$ to $y$.  A node of the path is adjacent
to at most two nodes of the path, because no edge joins two nodes of the
path that are not consecutive.  A node outside the path is adjacent to at
most three nodes of the path, because the path is shortest.  Hence the
degrees of the nodes of the path sum to less than $3n$.  The resistance of
an edge $\{u,v\}$ is $2\max\{d(u),d(v)\}\le2(d(u)+d(v))$, and every node of
the path is an end of at most two edges of the path.  Thus the resistances
along the path sum to less than $4\cdot3n=12n$.  It follows
that $R(x,y)<12n$ and $H(x,y)<12n^2$.
\end{proof}

An agent simulates one step of the lazy Metropolis walk by the following
procedure.  The procedure replaces the random choices of the step by the
entries of a list of integers, which we call a row.  Let $N\ge2$ be an
integer known to the agent.  A \emph{row} consists of the following
entries.
\begin{itemize}
\item A bit $\lambda$, which decides whether the agent stays.
\item A port number $p_d\in\{1,\dots,d\}$ for every degree $d$ with
$1\le d<N$.
\item An integer $U_{d,d'}\in\{1,\dots,d'\}$ for all degrees $d$ and $d'$ with
$1\le d,d'<N$.
\end{itemize}
Procedure \Transition\ takes two rounds (\cref{alg:transition}).  The agent
takes port $p_d$, where $d$ is the degree of its current node.  At the new
node, of degree $d'$, it uses the entry $U_{d,d'}$ to decide whether it stays
or returns.

\begin{algorithm}[ht]
\caption{Procedure \Transition}
\label{alg:transition}
\begin{algorithmic}[1]
\Require a row $(\lambda,(p_d),(U_{d,d'}))$; the agent is at a node of degree $d$
\If{$d=0$, $d\ge N$ or $\lambda=1$}
  \State stay for two rounds
\Else
  \State take port $p_d$; let $d'$ be the degree of the new node
  \If{$d'<N$ and $U_{d,d'}\le\min\{d,d'\}$}
    \State stay for one round
  \Else
    \State return by the port by which it entered
  \EndIf
\EndIf
\end{algorithmic}
\end{algorithm}

\begin{lemma}
\label{lem:transition}
Procedure \Transition\ takes exactly two rounds in any graph.  Suppose that
the graph has size at most $N$ and that all entries of the row are chosen
independently and uniformly at random.  Then the node of the agent after
the procedure is distributed as one step of the lazy Metropolis walk.
\end{lemma}
\begin{proof}
Each branch of the procedure takes two rounds.  In a graph of size at most $N$ all degrees are smaller than $N$.  The agent moves to a fixed
neighbour of degree $d'$ and stays there if $\lambda=0$, the port $p_d$ leads
to this neighbour, and $U_{d,d'}\le\min\{d,d'\}$.  The probability of these
three events is
$\frac12\cdot\frac1d\cdot\frac{\min\{d,d'\}}{d'}=\frac1{2\max\{d,d'\}}$, as
required by \eqref{eq:metropolis}.
\end{proof}

\subsection{Agents that start simultaneously}
\label{sec:sim}

Let $\cF$ be a finite family of connected port-labelled graphs, and let
$F$ be the number of its members.  We assume that both agents know $\cF$,
that every graph of $\cF$ has at least two nodes and has size at most $N$, and that the graph in which
the agents operate belongs to $\cF$.  Throughout this section, we let
$h=12N^2$ and $\alpha=\lceil\log(FN)\rceil$.  By \cref{lem:hit}, all hitting
times in the graphs of $\cF$ are smaller than $h$.

The algorithm is built from units called trials, which we now define.  A \emph{block} is one of the following
two instructions.  A \emph{waiting block} tells the agent to stay for $4h$
rounds.  A \emph{walking block} consists of $2h$ rows and tells the agent
to execute Procedure \Transition\ once with each of them.  Hence every block
takes $4h$ rounds.  Let $\kappa\ge1$ be an integer.  A \emph{key} is an
integer in $\{0,\dots,2^\kappa-1\}$.  A \emph{trial table} for $\kappa$
assigns a sequence of $8(\alpha+\kappa)$ blocks to every key.  In a
\emph{trial}, an agent executes the blocks that the table assigns to its
key (\cref{fig:trial}(a)).  A trial thus takes exactly $32h(\alpha+\kappa)$ rounds.

\begin{figure}[ht]
\centering
\begin{tikzpicture}[x=1cm,y=1cm,
  wb/.style={draw,fill=cbBlue!30,minimum width=1.3cm,minimum height=0.6cm,inner sep=0pt,font=\scriptsize},
  sb/.style={draw,fill=white,minimum width=1.3cm,minimum height=0.6cm,inner sep=0pt,font=\scriptsize}]
\foreach \t [count=\c from 0] in {wb,sb,sb,wb,wb,sb,wb,sb}
  \node[\t] (a\c) at (1.3*\c,1.0) {};
\foreach \t [count=\c from 0] in {wb,sb,wb,wb,sb,sb,sb,wb}
  \node[\t] (b\c) at (1.3*\c,0) {};
\foreach \c in {0,3,4,6} \node[font=\scriptsize] at (1.3*\c,1.0) {walk};
\foreach \c in {1,2,5,7} \node[font=\scriptsize] at (1.3*\c,1.0) {stay};
\foreach \c in {0,2,3,7} \node[font=\scriptsize] at (1.3*\c,0) {walk};
\foreach \c in {1,4,5,6} \node[font=\scriptsize] at (1.3*\c,0) {stay};
\node[anchor=east,font=\small] at (-0.85,1.0) {key $1$};
\node[anchor=east,font=\small] at (-0.85,0) {key $2$};
\foreach \c in {2,4,6,7}
  \draw[cbOrange,very thick] (1.3*\c-0.65,-0.42) rectangle (1.3*\c+0.65,1.42);
\draw[decorate,decoration={brace,amplitude=3pt}] (-0.65,1.5)--(0.65,1.5)
  node[midway,above=3pt,font=\scriptsize]{one block: $4h$ rounds};
\node[font=\small] at (4.55,-0.95) {(a) a trial: the blocks of two keys};
\begin{scope}[shift={(-0.65,-3.7)},x=0.58cm]
\foreach \u in {0,1,2,3}{
  \draw[fill=black!12] (4*\u,1.0) rectangle (4*\u+2,1.5);
  \draw[fill=cbBlue!30] (4*\u+2,1.0) rectangle (4*\u+4,1.5);
}
\foreach \u in {0,1,2}{
  \draw[fill=black!12] (5+4*\u,0) rectangle (5+4*\u+2,0.5);
  \draw[fill=cbBlue!30] (5+4*\u+2,0) rectangle (5+4*\u+4,0.5);
}
\foreach \r in {1,...,15} \draw[black!40] (\r,1.0)--(\r,1.5);
\foreach \r in {6,...,16} \draw[black!40] (\r,0)--(\r,0.5);
\draw[cbOrange,very thick] (9,-0.12) rectangle (10,1.62);
\draw[<->] (0,0.25)-- node[fill=white,inner sep=1pt,font=\scriptsize]{delay $\tau=5$} (5,0.25);
\node[anchor=east,font=\small] at (-0.4,1.25) {earlier agent};
\node[anchor=east,font=\small] at (-0.4,0.25) {later agent};
\foreach \r in {0,4,8,12,16} \node[font=\tiny] at (\r,1.72) {\r};
\node[font=\tiny,anchor=west] at (16.6,1.72) {rounds};
\draw[fill=black!12] (0,-0.9) rectangle (0.8,-0.6); \node[anchor=west,font=\scriptsize] at (0.9,-0.75) {stay};
\draw[fill=cbBlue!30] (3.2,-0.9) rectangle (4.0,-0.6); \node[anchor=west,font=\scriptsize] at (4.1,-0.75) {Procedure \Transition};
\node[font=\small] at (8.5,-1.6) {(b) padded transitions of two agents that start with a delay};
\end{scope}
\end{tikzpicture}
\caption{The two ways in which the algorithms use the labels.  (a) In a trial,
each agent executes a sequence of blocks that depends on its key.  In a
walking block the agent simulates $2h$ steps of the random walk, and in a
waiting block it stays.  The orange frames mark the blocks in which one
agent walks and the other agent stays.  In each of them the walking agent
reaches the other agent with probability at least $\frac12$.  The agents
start the trial in the same round, so their blocks are aligned.  (b) In
Algorithm \Tape, each agent repeats padded transitions of four rounds.  It
stays for two rounds and then executes Procedure \Transition.  Here the
delay is $5$.  The earlier agent is at one node from round $8$ to round $10$,
after two transitions.  The later agent is at one node from round $9$ to
round $11$, after one transition.  The two intervals overlap (orange frame).
Hence, if the two transitions lead the agents to the same node, then the
agents meet.}
\label{fig:trial}
\end{figure}

\begin{lemma}[One trial]
\label{lem:trial}
For every $\kappa\ge1$ there exists a trial table for $\kappa$ with the
following property.  Two agents that have distinct keys and start a trial
in the same round meet during the trial, in every graph of $\cF$ and from
every pair of distinct nodes.
\end{lemma}
\begin{proof}
Choose a random table.  Each block is a waiting block or a walking block
with probability $\frac12$, and all entries of all rows are uniform.  All
choices are independent.  Fix a graph of $\cF$, a pair of starting nodes
and a pair of distinct keys.  Consider one block index, and condition on
the earlier blocks.  We may assume that the agents are at different nodes
at the beginning of the block.  Otherwise they have met, at the end of the
previous round.  With probability $\frac12$ one agent has a waiting
block and the other agent has a walking block.  In this case the waiting
agent is a fixed target.  By \cref{lem:transition}, the walking agent makes
$2h$ steps of the lazy Metropolis walk.  By \cref{lem:hit}, the expected number of steps
until it reaches the target is smaller than $h$.  By Markov's inequality,
the number of steps exceeds $2h$ with probability at most $\frac12$.
Hence the agents fail to meet during the block with probability at most
$\frac34$.  The probability that they do not meet during the trial is at
most $(3/4)^{8(\alpha+\kappa)}$.

The number of choices of a graph, two nodes and two keys is at most
$FN^22^{2\kappa}\le2^{2(\alpha+\kappa)}$.  Since
$2^{2(\alpha+\kappa)}(3/4)^{8(\alpha+\kappa)}<1$, the probability that the
random table fails for some choice is smaller than $1$.  Hence some table
succeeds for all choices.
\end{proof}

The existence of a trial table is all that the agents need.  For every
$\kappa$, both agents test all trial tables for $\kappa$ in a fixed order
and use the first table that has the property of \cref{lem:trial}.  The
test of a table simulates it in all graphs of $\cF$.  Hence both agents
obtain the same table.

We say that a trial \emph{succeeds} if the agents meet during the trial.
\Cref{lem:trial} guarantees a success only if the keys of the agents are
different.  An agent
does not know the length of the other label, so it cannot choose $\kappa$
in advance.  Algorithm \Trials\ (\cref{alg:trials}) therefore executes
trials with $\kappa_i=2^i\alpha$ for $i=0,1,\dots$ until the meeting.  The
variable $i$ is the index of the current trial.  The variable
$\mathit{key}$ is the key of the agent in the current trial.  The key is the label of the agent if
the label is smaller than $2^{\kappa_i}$, and $0$ otherwise.

\begin{algorithm}[ht]
\caption{\Trials, for an agent with label $\ell$}
\label{alg:trials}
\begin{algorithmic}[1]
\Require the family $\cF$ and the bound $N$
\For{$i=0,1,2,\dots$ until the meeting}
  \State $\kappa_i\gets2^i\alpha$
  \State \textbf{if} $\ell<2^{\kappa_i}$ \textbf{then} $\mathit{key}\gets\ell$
         \textbf{else} $\mathit{key}\gets0$
  \State compute the first trial table for $\kappa_i$ with the property of
         \cref{lem:trial}
  \State execute the $8(\alpha+\kappa_i)$ blocks that the table assigns to
         the key $\mathit{key}$
\EndFor
\end{algorithmic}
\end{algorithm}

\begin{proposition}[Known family, simultaneous start]
\label{prop:known}
Algorithm \Trials\ guarantees rendezvous within
$O\bigl(h(\log F+\log N+b)\bigr)$ rounds, for agents with distinct labels
that start simultaneously in a graph of $\cF$.
\end{proposition}
\begin{proof}
The length of a trial does not depend on the key or on the graph.  Hence
both agents start every trial in the same round.  Let $i_0$ be the smallest
index with $\kappa_{i_0}\ge b$.  In trial $i_0$ the shorter label is smaller than
$2^{\kappa_{i_0}}$, so the agent with this label uses its label as the key.  The
label is positive.  The key of the other agent is either $0$ or the other
label.  In both cases the two keys are different.  If the agents are at the
same node at the beginning of trial $i_0$, then they have met already.
Otherwise, by \cref{lem:trial}, they meet during trial $i_0$.

By the choice of $i_0$ we have $\kappa_{i_0}\le2\max\{\alpha,b\}$.  The trials
$0,\dots,i_0$ take $32h\sum_{i\le i_0}(\alpha+\kappa_i)<128h\kappa_{i_0}$ rounds in
total, because $\alpha\le\kappa_i$ and $\sum_{i\le i_0}\kappa_i<2\kappa_{i_0}$.
\end{proof}

We now apply \cref{prop:known} to the family of all connected graphs of size at most $N$.  Such a graph has fewer than $N^2$ edges.  Name its nodes
$1,2,\dots$ in an arbitrary way.  Then each edge is described by the names
of its two ends and its two port numbers, which are four integers smaller
than $N+1$.  Hence
\begin{equation}
 F\le(N+1)^{4N^2+4}\qquad\text{and}\qquad\log F=O(N^2\log N).
 \label{eq:count}
\end{equation}
With $h=12N^2$, \cref{prop:known} gives the time $O(N^4\log N+N^2b)$, for
agents that know a bound $N\ge n$ and start simultaneously.  The algorithm of
\cref{sec:unknown} needs neither assumption.

The same method gives an exploration procedure for one agent.

\begin{remark}[Exploration by one agent]
\label{rem:explore}
There exists a sequence of $2h(2\alpha+1)$ rows with the following
property.  An agent that executes Procedure \Transition\ with these rows,
one after another, visits all nodes, in every graph of $\cF$ and from every
starting node.  Indeed, choose the rows at random and fix a graph of $\cF$,
a starting node and a target node.  Divide the rows into $2\alpha+1$ runs
of $2h$ rows.  By \cref{lem:hit} and Markov's inequality, each run reaches
the target with probability at least $\frac12$, whatever happened before.
Hence the agent does not visit the target with probability at most
$2^{-(2\alpha+1)}$.  There are at most $FN^2\le2^{2\alpha}$ choices of a
graph and two nodes.  Thus some sequence succeeds for all choices.  The
argument remains valid if $\alpha$ is replaced by a larger integer.  For the
family of all connected graphs of size at most $N$, the agent visits all
nodes within $4h(2\alpha+1)=O(N^4\log N)$ rounds.
\end{remark}

\subsection{Agents that start with an arbitrary delay}
\label{sec:delay}

If the agents start in different rounds, then their trials are not
aligned, and the guarantee of Algorithm \Trials\ is lost.  We now describe Algorithm \Tape,
which works for any delay.  The idea is that an agent no longer alternates
between waiting and walking.  Each agent walks all the time, and it repeats
one finite sequence of rows forever.  The sequences of two different labels
are chosen so that the two walks meet, wherever in their sequences the
agents are.

A \emph{tape} is a finite sequence of rows.  Every label $\ell$ is assigned
a tape, whose number of rows we denote by $M_\ell$.  A \emph{padded
transition} takes four rounds.  The agent stays for two rounds and then
executes Procedure \Transition.  An agent with label $\ell$ executes padded
transitions with the rows of its tape in cyclic order.  After the last row
it continues with the first row (\cref{alg:tape} and
\cref{fig:trial}(b)).  The variable $\mathit{row}$ is the
index of the current row.  The \emph{phase} of an agent is the value of
$\mathit{row}$ at a given time.  We also consider agents that start with a
phase other than $1$, which means that they begin with another row of their
tape.

\begin{algorithm}[ht]
\caption{\Tape, for an agent with label $\ell$}
\label{alg:tape}
\begin{algorithmic}[1]
\Require the family $\cF$ and the bound $N$
\State compute the tapes of the labels $1,\dots,\ell$ as in the proof of
       \cref{prop:cyclic}
\State $\mathit{row}\gets1$
\Loop\ until the meeting
  \State stay for two rounds
  \State execute Procedure \Transition\ with row $\mathit{row}$ of the tape of $\ell$
  \State \textbf{if} $\mathit{row}<M_\ell$ \textbf{then}
         $\mathit{row}\gets\mathit{row}+1$ \textbf{else} $\mathit{row}\gets1$
\EndLoop
\end{algorithmic}
\end{algorithm}

The analysis rests on three statements.  \Cref{lem:moving} bounds the probability
that a random walk avoids a moving target.  \Cref{lem:offset} shows that
padded transitions absorb a delay that is not a multiple of four.
\Cref{prop:cyclic} proves that suitable tapes exist.

\Cref{lem:moving} needs one more notion.  A \emph{stationary distribution} of a
random walk is a probability distribution on the nodes that is preserved by
one step of the walk.  For a symmetric walk on a connected graph it is the
uniform distribution.  The following lemma is due to Oliveira and
Peres~\cite{OP19}.

\begin{lemma}[Moving target~\cite{OP19}]
\label{lem:moving}
Consider the lazy Metropolis walk $X_0,X_1,\dots$ on a connected graph of
size at most $N$, started at any node.  Let $h$ be an upper bound on its
hitting times.  For any fixed sequence $y_0,\dots,y_t$ of nodes, the
probability that $X_{t'}\ne y_{t'}$ for all $0\le t'\le t$ is at most
$N\mathrm e^{-t/h}$.
\end{lemma}
\begin{proof}
Oliveira and Peres~\cite[Theorem~5 of the arXiv version, for reversible chains
with nonnegative spectrum]{OP19} prove the bound $(1-1/h)^t\le\mathrm
e^{-t/h}$ for a walk that starts from its stationary distribution.  Their
assumptions hold for every lazy symmetric walk.  The stationary
distribution is uniform, so it gives probability at least $1/N$ to the
starting node.  Hence the bound for a fixed starting node is at most $N$
times larger.
\end{proof}

\begin{lemma}[Delays]
\label{lem:offset}
Let $M$ be an integer.  Consider two tapes with the following property in a
given graph.  Let the two agents start simultaneously, from any pair of
nodes and with any pair of phases.  Then, for some $1\le t\le M$, both agents
are at the same node after each of them has executed Procedure \Transition\
$t$ times with the consecutive rows of its tape.  Under this
assumption, two agents that execute Algorithm \Tape\ with these tapes meet
within $4M+2$ rounds after the start of the later agent, for any delay.
\end{lemma}
\begin{proof}
Let $\tau$ be the delay, and let $t_0$ be the integer nearest to $\tau/4$.
Hence $|\tau-4t_0|\le2$.  Count the rounds from the start of the earlier
agent.  The earlier agent completes $t_0$ padded transitions by round $4t_0$.
We compare its transitions from that round on with the transitions of the
later agent from its start.  By the assumption, for some $1\le t\le M$ the two
agents are at the same node after $t$ further transitions each.  Every
padded transition begins with two rounds in which the agent stays.  Hence
the
earlier agent is at this node in the rounds $4(t_0+t)$, $4(t_0+t)+1$ and
$4(t_0+t)+2$.  The later agent is at this node in the rounds $\tau+4t$,
$\tau+4t+1$ and $\tau+4t+2$.  Since $|\tau-4t_0|\le2$, the two sets of rounds
have a common element.
\end{proof}

The proof of \cref{prop:cyclic} uses the Lov\'asz local lemma~\cite{EL75} in the following
form~\cite{AS16}.  Consider a finite set of events.  Suppose that every event $A$ of the set
is independent of the collection of all events that are neither $A$ nor in a
set $\Gamma(A)$.
We say that $A$ is \emph{dependent} on the events of $\Gamma(A)$.
Suppose also that every event $A$ is assigned a number $\xi(A)\in[0,1)$
such that the probability of $A$ is at most
$\xi(A)\prod_{A'\in\Gamma(A)}(1-\xi(A'))$.  Then with positive probability
none of the events occurs.

\begin{proposition}[Known family, arbitrary delay]
\label{prop:cyclic}
The tapes can be chosen so that Algorithm \Tape\ guarantees rendezvous
within $O\bigl(h(\log F+\log N+b)\bigr)$ rounds after the start of the later
agent, for agents with distinct labels in a graph of $\cF$ and for any
delay.
\end{proposition}
\begin{proof}
Let $\gamma=\lceil\log(2FN^3(h+1)^2)\rceil$.  The tape of a label $\ell$ has
$M_\ell=64h(\gamma+\len{\ell})$ rows.  We define the tapes by induction on the
label.  The tape of label $1$ is arbitrary.  Let $\ell\ge2$, and suppose
that the tapes of all smaller labels are defined.  We claim that there
exists a tape for $\ell$ with the following property.  For every label
$\ell'<\ell$, the tape of $\ell$ and the tape of $\ell'$ satisfy the
assumption of \cref{lem:offset} with $M=M_{\ell'}$, in every graph of
$\cF$.  The agents find the first such tape by exhaustive search.  The claim
and \cref{lem:offset} prove the proposition.  Indeed, the agents meet within
$4M_{\ell'}+2$ rounds, where $\ell'$ is the smaller label, and
$\len{\ell'}=b$.

To prove the claim, we choose the $M_\ell$ rows of the new tape
independently and uniformly at random.  An event is described by a label
$\ell'<\ell$, a graph of $\cF$, two nodes, a phase of the tape of $\ell'$
and a phase of the new tape.  The event occurs if the two agents, started
simultaneously with these parameters, are at different nodes after each of
the first $M_{\ell'}$ transitions.  The agent with the tape of $\ell'$ is a
fixed moving target.  Since $M_{\ell'}\le M_\ell$, the event depends on
$M_{\ell'}$ distinct rows of the new tape.  We apply \cref{lem:moving} to
the agent with the new tape.  The nodes $y_1,y_2,\dots$ are the nodes of the
other agent after its transitions, and $y_0$ is any node other than the
starting node of the agent with the new tape.  By \cref{lem:transition} and
\cref{lem:moving}, the probability of the event is at most $N\mathrm e^{-M_{\ell'}/h}$.

Two events are dependent only if their sets of rows intersect.  An event
for the label $\ell'$ depends on $M_{\ell'}$ consecutive rows.  It is
dependent on at most $(M_{\ell'}+M_{\ell''})\cdot FN^2M_{\ell''}$ events for a
label $\ell''$.  We assign the number
$\xi_{\ell''}=\mathrm e^{-M_{\ell''}/(2h)}$ to every event for the label
$\ell''$.  There are fewer than $2^i$ labels of length $i$, and
$2^{\gamma}\ge2FN^3(h+1)^2$.  Using the inequality
$2^{\gamma+i}(\gamma+i)^2\le\mathrm e^{2(\gamma+i)}$, we obtain
\[
 FN^2\sum_{\ell''\ge1}M_{\ell''}^2\xi_{\ell''}
 \le 4096\,FN^2h^2\sum_{i\ge1}2^i(\gamma+i)^2\mathrm e^{-32(\gamma+i)}
 \le 2048\sum_{i\ge1}\mathrm e^{-30(\gamma+i)}<\tfrac1{64}.
\]
Since $M_{\ell''}\ge h$, the sum of the numbers $\xi_{\ell''}$ over the events that are
dependent on a given event for $\ell'$ is at most
$M_{\ell'}/(64h)+\frac1{64}\le M_{\ell'}/(32h)$.  Since $\ln(1-\xi)\ge-2\xi$ for
$0\le\xi\le\frac12$, the product required by the local lemma is at least
\[
 \mathrm e^{-M_{\ell'}/(2h)}\cdot\mathrm e^{-M_{\ell'}/(16h)}
 \ge N\mathrm e^{-M_{\ell'}/h}.
\]
The last inequality is equivalent to $M_{\ell'}/h\ge\frac{16}7\ln N$, which
holds because $M_{\ell'}/h\ge64\gamma$.  By the local lemma, with positive
probability no event occurs.  Hence the required tape exists.
\end{proof}

\begin{proof}[Proof of \cref{cor:fourcover}]
A four-sheet cover of $K_q$ is given by
one permutation of four sheets for each of the $s$ edges of $K_q$.  Hence
the number of these covers is at most $24^s$, and $\log F=O(n^2)$.  By
\cref{prop:cyclic} with $N=n$, the time is $O(n^4+n^2b)$.  The matching
lower bound is \cref{cor:quartic-fourcover}.
\end{proof}

For the family of all connected graphs of size at most $N$,
\cref{prop:cyclic} and the bound \eqref{eq:count} give the time
$O(N^4\log N+N^2b)$ for any delay, for agents that know $N$.  We now remove
the assumption that the agents know a bound on the size.

\subsection{Agents that know nothing}
\label{sec:unknown}

\begin{figure}[ht]
\centering
\begin{tikzpicture}[x=1cm,y=1cm]
\draw[fill=black!12] (0,3.4) rectangle (6.2,4.2);
\node[font=\scriptsize] at (3.1,3.8) {waiting period: the agent stays};
\draw[fill=cbBlue!45] (6.2,3.4) rectangle (7.5,4.2);
\draw[fill=cbBlue!25] (7.5,3.4) rectangle (11.2,4.2);
\draw[fill=cbBlue!8] (11.2,3.4) rectangle (12.4,4.2);
\node[font=\scriptsize] at (6.85,3.93) {explore};
\node[font=\scriptsize] at (9.35,3.93) {frame: $2T_i$ free rounds};
\node[font=\scriptsize] at (11.8,3.93) {stay};
\foreach \x in {6.3,6.75,7.2,7.65,8.1,8.55,9.0,9.45,9.9,10.35,10.8,11.25,11.7,12.15}
  \fill[cbOrange] (\x,3.4) rectangle (\x+0.06,3.62);
\foreach \x in {6.5,8.75,11.0}
  \fill[cbOrange] (\x,3.4) rectangle (\x+0.18,3.62);
\draw (6.2,3.4) rectangle (12.4,4.2);
\draw[decorate,decoration={brace,amplitude=4pt}] (0,4.3)--(6.2,4.3)
  node[midway,above=4pt,font=\scriptsize]{$R_i$ rounds};
\draw[decorate,decoration={brace,amplitude=4pt}] (6.2,4.3)--(12.4,4.3)
  node[midway,above=4pt,font=\scriptsize]{active period: $R_i$ rounds};
\node[font=\small] at (6.2,2.95) {(a) stage $i$ of one agent};
\begin{scope}[shift={(0,1.3)}]
\draw[fill=cbBlue!12] (0,0.7) rectangle (5.6,1.2);
\foreach \x in {0.5,1.4,2.3,3.2,4.1,5.0}
  \draw[cbOrange,line width=2.4pt] (\x,0.7)--(\x,1.2);
\draw[fill=black!12] (1.1,0) rectangle (3.9,0.5);
\draw[fill=cbBlue!12] (3.9,0) rectangle (5.6,0.5);
\node[font=\scriptsize] at (3.15,0.25) {waiting};
\draw[cbOrange,very thick] (1.25,-0.1) rectangle (2.45,1.3);
\node[anchor=east,font=\scriptsize] at (-0.15,0.95) {earlier agent};
\node[anchor=east,font=\scriptsize] at (-0.15,0.25) {later agent};
\node[font=\small] at (2.8,-0.5) {(b) large delay};
\end{scope}
\begin{scope}[shift={(7.4,1.3)}]
\draw[fill=cbBlue!25] (0,0.7) rectangle (4.2,1.2);
\draw[fill=cbBlue!25] (0.8,0) rectangle (5.0,0.5);
\node[font=\scriptsize] at (2.1,0.95) {frame};
\node[font=\scriptsize] at (2.9,0.25) {frame};
\draw[cbOrange,very thick] (0.8,-0.1) rectangle (4.2,1.3);
\draw[<->] (0,0.25)-- node[below=1pt,font=\scriptsize]{$\tau$} (0.8,0.25);
\node[font=\small] at (2.5,-0.5) {(c) small delay};
\end{scope}
\end{tikzpicture}
\caption{Algorithm \Windows\ (not to scale).  (a) Stage $i$ consists of a
waiting period and an active period.  In the active period the agent first
explores the graph for the bound $N_i$.  Then it executes a frame, in which
it simulates a random walk in $2T_i$ free rounds.  The orange marks at the bottom
are the inserted closed explorations, which interrupt the other
instructions.  The wide marks are explorations for a large bound, and the
narrow marks are the shorter and more frequent explorations for smaller
bounds.  (b) If the
delay is large, then some window (orange frame) lies in a waiting period of
one agent and contains a complete closed exploration of the other agent.
The exploring agent visits the node of the waiting agent.  (c) If the delay
$\tau$ is small, then both agents execute the frame of the same stage, and
the two frames overlap (orange frame).  The rows of the two labels are
chosen so that the two walks meet in the overlap.}
\label{fig:windows}
\end{figure}
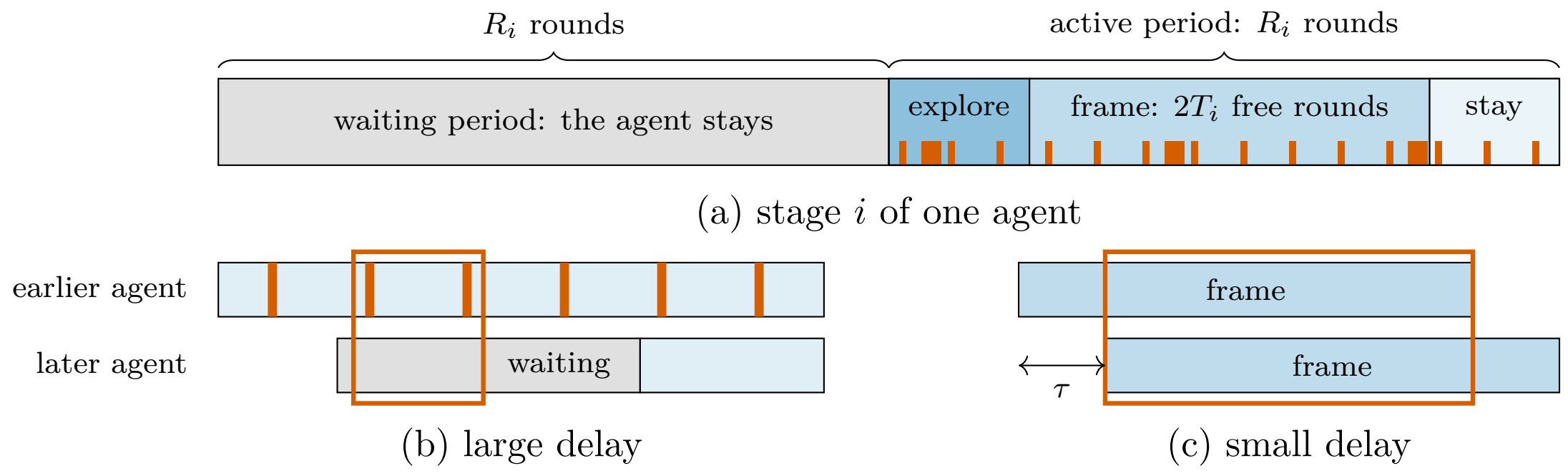

We now prove \cref{thm:upper}.  The agents know no bound on the size of the
graph and no bound on the labels, and an adversary chooses the delay.  The
algorithms of \cref{sec:sim,sec:delay} do not apply, because their tables
are computed for a known family of graphs.  The main idea of Algorithm
\Windows\ is the following.  The algorithm works in stages, and stage $i$
is designed for the graphs of size at most $N_i=2^{i+1}$.  Every stage
begins with a long period in which the agent stays at its node.  In the
rest of the stage, the agent explores the graph again and again, and
between the explorations it makes a random walk as in \cref{sec:delay}.  If
the delay is large, then at some time one agent stays while the other agent
completes an exploration.  Hence the agents meet, and the labels play no
role.  If the delay is small, then both agents are in the same stage, and
their executions are shifted by a bounded number of rounds.  In this case
the two walks meet (\cref{fig:windows}).

Two difficulties arise.  First, an agent in stage $i$ does not know which
of the bounds $N_0,\dots,N_i$ is the first that is at least $n$.  It has to
explore often enough for each of them, and explorations of full length for
all bounds would take too many rounds.  We use the fact that an agent that
has explored the graph has seen its maximum degree.  A graph of small
maximum degree is described by few bits, so it has a short exploration.
Hence the explorations for the bounds far above the observed degree are
short, and the bounds below the observed degree need no exploration.
Second, the rows of the random walk have to succeed for all sizes up to
$N_i$ and for all labels, although the lengths of the labels are not known.
We choose the rows label by label, in the increasing order of the labels.
The rows of a label are tested only against the smaller labels that are
short enough for the size of the graph.

Closed explorations inserted at summable rates, and the analysis of
arbitrary windows of rounds, also appear in the exploration sequences of
Xin~\cite{Xin26} for one agent.  The new ingredients here are the degree
bound and the frames.

We first define explorations that return the agent to its node
(\cref{lem:closed}).  Next, we insert such explorations into the execution
of an agent (\cref{lem:windows}), and we define the frames in which the
agents walk (\cref{lem:frames}).  Then we present the algorithm and prove
the theorem.

For an integer $\nu\ge2$, let
\[
 h(\nu)=12\nu^2\qquad\text{and}\qquad
 A(\nu)=\bigl\lceil\log\bigl(\nu(\nu+1)^{4\nu^2+4}\bigr)\bigr\rceil .
\]
By \cref{lem:hit} and by \eqref{eq:count}, the hitting times in a graph of
size at most $\nu$ are smaller than $h(\nu)$, and the number of connected
graphs of size at most $\nu$ is at most $2^{A(\nu)}/\nu$.  For $g\ge0$ let
$N_g=2^{g+1}$, $h_g=h(N_g)$ and $\alpha_g=A(N_g)$.  We say that $D$ is a
\emph{degree bound for $g$} if $D=N_j$ for some $0\le j\le g$.  For such a
$D$, let
\[
 \alpha_{g,D}=\bigl\lceil\log\bigl(N_g(N_g+1)^{4N_gD+4}\bigr)\bigr\rceil .
\]
Consider the family of the connected graphs of size at most $N_g$ in which
every degree is at most $D$.  Such a graph has at most $N_gD/2$ edges.
Hence, by the encoding that gives \eqref{eq:count}, the family has at most
$(N_g+1)^{4N_gD+4}$ members.  By \cref{rem:explore}, applied to this family
with $\alpha_{g,D}$ in the role of $\alpha$, there is a sequence of
$2h_g(2\alpha_{g,D}+1)$ rows for the bound $N_g$ with which an agent visits all nodes of every
graph of the family.  The agents use the first such sequence in a fixed
order of all sequences of rows, so all agents use the same sequence.  In a
\emph{closed exploration for $(g,D)$}, the agent
first executes Procedure \Transition\ with these rows, one after another.
For every row, it remembers whether the procedure moved it to a neighbour
and, if so, the port by which it entered.  Then the agent considers the
rows in the reverse order.  For a row that moved it, the agent takes the
remembered port and stays for one round.  For any other row, it stays for
two rounds.  We denote by $u_{g,D}=8h_g(2\alpha_{g,D}+1)$ the number of
rounds of a closed exploration for $(g,D)$.  For $D=N_g$ the family
consists of all connected graphs of size at most $N_g$, and
$\alpha_{g,N_g}=\alpha_g$.  We write $u_g$ for $u_{g,N_g}$.

\begin{lemma}[Closed exploration]
\label{lem:closed}
Let $D$ be a degree bound for $g$.
\begin{enumerate}[label=(\roman*)]
\item In any graph, a closed exploration for $(g,D)$ takes exactly
$u_{g,D}$ rounds and ends at the node at which it starts.
\item In a graph of size at most $N_g$ and of maximum degree at most $D$,
the agent visits all nodes during the exploration.
\item We have $4u_{g,D}\le u_{g+1,D}$, $u_{g,D}\le(2D/N_g)\,u_g$ and
$4u_g\le u_{g+1}\le36u_g$.
\end{enumerate}
\end{lemma}
\begin{proof}
Procedure \Transition\ takes two rounds.  It either leaves the agent at its
node or moves it along one edge.  Taking the port by which the agent
entered undoes such a move.  Hence the second half of the exploration
undoes the transitions of the first half in the reverse order, which
proves (i).  Claim (ii) is \cref{rem:explore}.  For (iii), recall that
$h_{g+1}=4h_g$.  The first inequality holds because
$\alpha_{g,D}\le\alpha_{g+1,D}$.  For the second inequality, we have
$2\alpha_{g,D}+1\le16N_gD\log(N_g+1)$ and
$2\alpha_g+1\ge8N_g^2\log(N_g+1)$, because $N_g\ge2$ and $D\ge2$.  The last
inequalities follow from $\alpha_g\le\alpha_{g+1}\le9\alpha_g$, which holds
because $\log(2N+1)\le2\log(N+1)$.
\end{proof}

A \emph{schedule} is an infinite sequence of rounds, together with the
instructions that an agent follows in these rounds.  Let $Z$ be a schedule,
let $D=N_j$ be a degree bound for $g$, and let
\[
 \omega_k=\frac1{16(k+1)^2}\qquad\text{for }k\ge0 .
\]
Thus $\sum_k\omega_k<\frac18$.  To \emph{insert closed explorations for
$(g,D)$} into $Z$ means to form the following schedule.  The agent executes
a closed exploration for $(g,D)$, then the next $u_{g,D}/\omega_{g-j}$
rounds of $Z$, then again a closed exploration for $(g,D)$, and so on.  An
inserted exploration may begin in the middle of Procedure \Transition.  The
agent remembers what it needs in order to continue.  By
\cref{lem:closed}(i), the agent is at the same node before and after an
inserted exploration.  Hence, in the rounds of $Z$, it is at the same nodes
as without the insertion.

Let $Y$ be a schedule and let $D=N_j$ be a degree bound for $i$.  We denote
by $Y^{(i,D)}$ the schedule obtained from $Y$ by inserting closed
explorations for $(i,D)$, then for $(i-1,D)$, and so on down to $(j,D)$,
each time into the schedule obtained so far.  A closed exploration for
$(g,D)$ is thus interrupted by closed explorations for smaller indices.  We
say that some consecutive rounds of $Y^{(i,D)}$ contain a \emph{complete}
closed exploration for $(g,D)$ if they contain all rounds of one closed
exploration for $(g,D)$, together with the rounds inserted between them.

\begin{lemma}[Windows]
\label{lem:windows}
Let $Y$ be a schedule, let $D=N_j$ be a degree bound for $i$, and let
$W_g=512u_g$ for $g\ge0$.
\begin{enumerate}[label=(\roman*)]
\item For every $g$ with $j\le g\le i$, any $W_g$ consecutive rounds of
$Y^{(i,D)}$ contain a complete closed exploration for $(g,D)$.
\item For every $x\ge1$, the first $2(x+4u_i/3)$ rounds of $Y^{(i,D)}$
contain the first $x$ rounds of $Y$.
\end{enumerate}
\end{lemma}
\begin{proof}
Let $Z'$ be obtained from a schedule $Z$ by inserting closed explorations
for $(g',D)$, and let $\omega=\omega_{g'-j}$.  Consider $x$ consecutive
rounds of $Z$, together with the rounds inserted between them and directly
before them.  At most $(x-1)\omega/u_{g',D}+1$ closed explorations are
inserted there.  Hence these rounds form at most
$(1+\omega)x+u_{g',D}\le(1+\omega)(x+u_{g',D})$ consecutive rounds of $Z'$.

Let $j\le g\le i$, and let $Y_g$ be the schedule obtained from $Y$ by the
insertions for $i,i-1,\dots,g$.  We apply the bound of the first paragraph
for $g'=g-1,\dots,j$.  The product of the factors $1+\omega_k$ is smaller
than $\mathrm e^{1/8}<2$, and $\sum_{g'<g}u_{g',D}\le u_{g,D}/3$ by
\cref{lem:closed}(iii).  Hence $x$ consecutive rounds of $Y_g$, together
with the rounds inserted between them and directly before them, form at
most $2(x+u_{g,D}/3)$ consecutive rounds of $Y^{(i,D)}$.  In particular,
any $2(1+u_{g,D}/3)$ consecutive rounds of $Y^{(i,D)}$ contain a round of
$Y_g$.

Let $u=u_{g,D}$ and $\omega=\omega_{g-j}$.  The schedule $Y_g$ alternates
between a closed exploration for $(g,D)$, which has $u$ rounds, and
$u/\omega$ other rounds.  Hence any $2u+u/\omega$ consecutive rounds of
$Y_g$ contain all rounds of one closed exploration for $(g,D)$.  Consider
now $2(1+u/3)+2(2u+u/\omega+u/3)$ consecutive rounds of $Y^{(i,D)}$.  They
contain a round of $Y_g$ among the first $2(1+u/3)$ of them, and then the
next $2u+u/\omega$ rounds of $Y_g$ with all rounds inserted between them.
Thus they contain a complete closed exploration for $(g,D)$.  Their number
is at most $6u+2u/\omega\le4u/\omega$, because $1/\omega\ge16$.  Let
$k=g-j$, so that $D/N_g=2^{-k}$.  By \cref{lem:closed}(iii),
\[
 \frac{4u}{\omega}=64(k+1)^2u_{g,D}\le128(k+1)^22^{-k}u_g\le288u_g\le W_g ,
\]
because $(k+1)^22^{-k}\le\frac94$ for every $k\ge0$.  Hence (i) holds.

For (ii), we apply the bound of the first paragraph for $g'=i,\dots,j$ to
the first $x$ rounds of $Y$.  They are among the first
$2(x+\sum_{g'\le i}u_{g',D})\le2(x+4u_{i,D}/3)$ rounds of $Y^{(i,D)}$, and
$u_{i,D}\le u_i$.
\end{proof}

The inequality in the proof of (i) is the place where the degree bound is
used.  The explorations for $(g,D)$ receive only the fraction
$\omega_{g-j}$ of the rounds.  They are nevertheless frequent, because
they are short: $u_{g,D}\le2^{1-(g-j)}u_g$ by \cref{lem:closed}(iii).

We now define the frames.  Let $B\ge2$ be an integer and let $T\ge2$ be an
even integer.  A \emph{frame} for $B$ and $T$ is a finite sequence of
rounds with the following instructions.  Exactly $2T$ of the rounds are
called \emph{free}.  In the other rounds, the agent executes closed
explorations that are inserted between the free rounds.  A \emph{frame
table} assigns a sequence of $T$ rows for the bound $B$ to every label.  An
agent \emph{executes the frame} as follows.  In the free rounds, it
executes Procedure \Transition\ with the $T$ rows of its label, one after
another.  In the other rounds, it executes the closed explorations.

\begin{lemma}[Frames]
\label{lem:frames}
For every frame for $B$ and $T$ there exists a frame table with the
following property.  Consider a graph of size $\nu$ with $2\le\nu\le B$,
and two agents with distinct labels.  Let $b$ be the length of the shorter
label, and suppose that
\[
 T\ge256\,h(\nu)\bigl(A(\nu)+b\bigr).
\]
Suppose that one agent starts to execute the frame $\tau$ rounds before the
other agent, for some $0\le\tau\le T$, and that the agents start from any
two nodes.  Then the agents meet in a round in which both of them execute
the frame.
\end{lemma}
\begin{proof}
We define the rows label by label, in the increasing order of the labels.
The rows of label $1$ are arbitrary.  Let $\ell\ge2$, and suppose that the
rows of all smaller labels are defined.  A \emph{configuration} for $\ell$
consists of an integer $\nu$ with $2\le\nu\le B$, a connected graph of size
$\nu$, a label $\ell'<\ell$ with $T\ge256\,h(\nu)(A(\nu)+\len{\ell'})$,
two nodes, an integer $\tau$ with $0\le\tau\le T$, and the choice of the
agent that starts earlier.  The agents of the configuration have the labels
$\ell$ and $\ell'$, start to execute the frame from the two nodes, and one
of them starts $\tau$ rounds before the other.  We say that the rows of
$\ell$ \emph{fail} for the configuration if these agents do not meet in any
round in which both of them execute the frame.  We claim that some $T$ rows
fail for no configuration.  The number of configurations is finite, so an
agent finds the first such sequence of rows by exhaustive search.  The claim
proves the lemma.  Indeed, of two distinct labels, the smaller label is not
longer than the other label, so its length is $b$.  Hence the two agents of
the lemma form a configuration for the larger label.

The frame is fixed before the rows are chosen: the positions of its free
rounds and the closed explorations in its other rounds do not depend on the
rows.  To prove the claim, we choose the $T$ rows of $\ell$ independently and
uniformly at random.  Fix a configuration, and let $h=h(\nu)$.  Let $X_0$
be the node at which the agent with label $\ell$ starts the frame, and let
$X_t$ be its node after its transition number $t$.  The closed explorations
of the frame return the agent to its node.  Hence, by \cref{lem:transition},
the nodes $X_0,\dots,X_T$ are consecutive steps of the lazy Metropolis
walk.  The rows of $\ell'$ are fixed.  Hence the node of the agent with
label $\ell'$ in every round is fixed.  Here we let both agents continue to
execute the frame after a meeting, which does not change whether they meet.

Consider first the case in which the agent with label $\ell$ does not start
earlier.  It completes its first $T/2$ transitions while $T$ free rounds of
its frame remain.  The other agent started at most $T$ rounds earlier and
executes the same frame.  Hence it still executes its frame in these
rounds.  For $1\le t\le T/2$, let $y_t$ be the node of the other agent at
the end of the round in which the agent with label $\ell$ completes its
transition number $t$, and let $y_0$ be any node other than $X_0$.  By
\cref{lem:moving}, the probability that $X_t\ne y_t$ for all $t\le T/2$ is
at most $\nu\,\mathrm e^{-T/(2h)}$.

Consider now the case in which the agent with label $\ell$ starts earlier.
It completes each of its last $T/2$ transitions at least $T+2$ rounds after
its start.  Hence the other agent has started, and it finishes its frame
later.  We condition on the first $T/2$ rows of $\ell$, which determine
$X_{T/2}$.  For $1\le t\le T/2$, let $y_t$ be the node of the other agent
at the end of the round in which the agent with label $\ell$ completes its
transition number $T/2+t$, and let $y_0$ be any node other than $X_{T/2}$.
By \cref{lem:moving}, applied to the walk started at $X_{T/2}$, the
probability that $X_{T/2+t}\ne y_t$ for all $t\le T/2$ is again at most
$\nu\,\mathrm e^{-T/(2h)}$.  In both cases, the rows fail for the
configuration with at most this probability.

It remains to count the configurations.  Fix $\nu$ and let $\mu=T/h(\nu)$.
If some configuration has the size $\nu$, then $\mu\ge256(A(\nu)+1)$.  The
labels $\ell'$ of the configurations of size $\nu$ have length at most
$\mu/256$, so their number is smaller than $2^{\mu/256}$.  There are at
most $2^{A(\nu)}/\nu$ graphs, at most $\nu^2$ pairs of nodes, and $2(T+1)$
choices of $\tau$ and of the agent that starts earlier.  We have
$T+1\le2T=24\nu^2\mu$ and $48\nu^4\le2^{A(\nu)}$.  Hence the expected
number of configurations of size $\nu$ for which the rows fail is at most
\[
 48\nu^4\mu\,2^{A(\nu)}2^{\mu/256}\mathrm e^{-\mu/2}
 \le\mu\,2^{2A(\nu)+\mu/256-\mu/2}\le2^{-\mu/4}\le2^{-\nu}.
\]
The second inequality holds because $2A(\nu)\le\mu/128$ and
$\log\mu\le\mu/8$, and the last inequality holds because $\mu\ge4\nu$.
Summing over $\nu\ge2$, the expected number of configurations for which the
rows fail is at most $\frac12$.  Hence some $T$ rows fail for no
configuration.
\end{proof}

The table of \cref{lem:frames} differs from the trial tables of
\cref{sec:sim} in one respect.  A trial table is computed for keys of a
given length.  The frame table serves all labels at once.  A pair of labels is covered by
the guarantee only if the smaller label is short enough for the given $T$.  An agent
with label $\ell$ computes only the rows of the labels $1,\dots,\ell$.

We now describe Algorithm \Windows\ (\cref{alg:windows}).  For $i\ge0$ let
\[
 R_i=8W_i=4096\,u_i\qquad\text{and}\qquad T_i=R_i/16=256\,u_i .
\]
Stage $i$ consists of a \emph{waiting period} and an \emph{active period},
of $R_i$ rounds each.  In the waiting period the agent stays.  At the
beginning of the active period, the agent fixes the degree bound $D$ of the
stage.  It is the smallest of the numbers $N_0,\dots,N_i$ that is at least
the largest degree that the agent has observed so far, and it is $N_i$ if
this degree is larger than $N_i$.  In the active period the agent executes
the first $R_i$ rounds of the schedule $Y_i^{(i,D)}$, where the schedule
$Y_i$ is defined as follows.  The agent first executes a closed exploration
for $(i,N_i)$, which takes $u_i$ rounds.  Then it has $2T_i$ free rounds,
in which it executes Procedure \Transition\ with $T_i$ rows.  Afterwards it
stays forever.  The rounds of $Y_i^{(i,D)}$ from the first free round to
the last free round form a frame for $N_i$ and $T_i$, which depends only on
$i$ and $D$.  The $T_i$ rows of the agent are those that the frame table of
\cref{lem:frames} for this frame assigns to its label.  In the pseudocode,
the variable $i$ is the index of the current stage, and the variable $D$ is
its degree bound.

\begin{algorithm}[ht]
\caption{\Windows, for an agent with label $\ell$}
\label{alg:windows}
\begin{algorithmic}[1]
\For{$i=0,1,2,\dots$ until the meeting}
  \State stay for $R_i$ rounds \Comment{waiting period}
  \State $D\gets$ the smallest of $N_0,\dots,N_i$ that is at least the
         largest degree observed so far, or $N_i$ if there is none
  \State compute the rows of $\ell$ in the frame table for the frame of
         $Y_i^{(i,D)}$
  \State execute the first $R_i$ rounds of the schedule $Y_i^{(i,D)}$
         \Comment{active period}
\EndFor
\end{algorithmic}
\end{algorithm}

The active period contains the closed exploration for $(i,N_i)$ and the
whole frame.  Indeed, these are the first $u_i+2T_i$ rounds of $Y_i$.  By
\cref{lem:windows}(ii), they are among the first
$2(u_i+2T_i+4u_i/3)=R_i/4+14u_i/3\le R_i$ rounds of $Y_i^{(i,D)}$.  The
number of rounds of a stage depends neither on the graph nor on the label
nor on the degree bound.  By \cref{lem:closed}(iii), we have
$4R_i\le R_{i+1}\le36R_i$.  It follows that the stages $0,\dots,i$ take
$2\sum_{i'\le i}R_{i'}\le\frac83R_i$ rounds.

\begin{proof}[Proof of \cref{thm:upper}]
Let $g_*$ be the smallest index with $N_{g_*}\ge n$, and let $g=g_*+1$.
Then $N_g<4n$.  Hence $h_g=O(n^2)$ and $u_g=O(n^4\log n)$.  Let $R=R_g$.
Let $D_*$ be the smallest of the numbers $N_0,N_1,\dots$ that is at least
the maximum degree of the graph.  The maximum degree is smaller than $n$,
so $D_*=N_j$ for some $j\le g_*$.

We first show that each agent uses the degree bound $D_*$ in every stage
with index at least $g$.  Indeed, in the active period of stage $g_*$ the
agent executes a complete closed exploration for $(g_*,N_{g_*})$.  By
\cref{lem:closed}(ii), it visits all nodes, because $n\le N_{g_*}$ and all
degrees are smaller than $N_{g_*}$.  Hence, from the end of its stage $g_*$
on, the largest degree that the agent has observed is the maximum degree of
the graph.

We count the rounds from the start of the later agent, and we let
$\tau\ge0$ be the delay.  For integers $x<y$, the \emph{window} $[x,y)$
consists of the rounds $x+1,\dots,y$.  Let $s$ be the number of rounds of
the stages $0,\dots,g-1$.  Thus the later agent is in the waiting period of
stage $g$ during $[s,s+R)$, in the active period of stage $g$ during
$[s+R,s+2R)$, and in the waiting period of stage $g+1$ during
$[s+2R,s+2R+R_{g+1})$.

We use the following fact.  Let $g'\ge g$, and consider a window of
$W_{g'}$ rounds.  Suppose that, during the whole window, one agent is in a
waiting period and the other agent is in the active period of a stage with
index at least $g'$.  Then the agents meet in the window.  Indeed, the
second agent uses the degree bound $D_*$.  By \cref{lem:windows}(i), it
executes a complete closed exploration for $(g',D_*)$ in the window.  By
\cref{lem:closed}(ii), it visits all nodes, because $n\le N_{g'}$ and the
maximum degree is at most $D_*$.

Consider first the case $\tau\ge2R$.  At the end of round $s$, the earlier
agent has executed at least $s+2R$ rounds.  Hence it has finished stage $g$, and each of its rounds after round $s$
belongs to a stage with index at least $g+1$.  Thus every active period of
the earlier agent after round $s$ belongs to such a stage, and each of its
periods that begins after round $s$ has at least $R_{g+1}\ge4R$ rounds.  Recall that $R=8W_g$ and
$R_{g+1}=8W_{g+1}\ge4R$.  We distinguish three cases.
\begin{itemize}
\item The earlier agent does not change its period during $[s,s+2R)$.  If it
is in an active period, then we apply the fact to the window $[s,s+W_g)$.
If it is in a waiting period, then we apply the fact to the window
$[s+R,s+R+W_g)$, in which the later agent is active.
\item The earlier agent changes from a waiting period to an active period at
the end of some round $\theta$ with $s<\theta<s+2R$.  If
$\theta\le s+R-W_g$, we apply the fact to the window $[s+R-W_g,s+R)$.  If
$\theta\ge s+R+W_g$, we apply the fact to the window $[s+R,s+R+W_g)$.
Otherwise the earlier agent is active during $[\theta,\theta+R_{g+1})$.
This window contains the window $[s+2R,s+2R+W_{g+1})$, because
$\theta<s+2R$ and $R_{g+1}\ge R+W_g+W_{g+1}$.  The later agent is in the
waiting period of stage $g+1$ in these rounds.  We apply the fact with
$g'=g+1$.
\item The earlier agent changes from an active period to a waiting period at
the end of some round $\theta$ with $s<\theta<s+2R$.  If $\theta\ge s+W_g$,
we apply the fact to the window $[s,s+W_g)$.  Otherwise the earlier agent
is in a waiting period during $[\theta,\theta+4R)$, which contains the
window $[s+R,s+R+W_g)$.  We apply the fact to this window.
\end{itemize}
In all cases the agents meet by round $s+2R+W_{g+1}$, which is
$O(R_{g+1})=O(R)$.

Consider now the case $\tau<2R$.  Let $i$ be the smallest index $i\ge g$
with
\[
 T_i\ge\max\bigl\{2R,\;256\,h(n)\bigl(A(n)+b\bigr)\bigr\}.
\]
In stage $i$ both agents use the degree bound $D_*$.  Hence they execute
the same frame for $N_i$ and $T_i$, and the earlier agent starts to execute
it $\tau<T_i$ rounds before the later agent.  The graph has size
$n\le N_i$.  By \cref{lem:frames}, the agents meet by the end of stage $i$
of the later agent, which is within $\frac83R_i$ rounds.  We have
$T_g=R/16<2R$, so $i>g$.  By the choice of $i$, the displayed inequality
fails for $i-1$.  Since $R_i\le36R_{i-1}=576\,T_{i-1}$, we obtain
$R_i=O\bigl(R+h(n)(A(n)+b)\bigr)$.

In both cases the time is $O\bigl(u_g+h(n)(A(n)+b)\bigr)$.  Since
$h(n)=12n^2$ and $A(n)=O(n^2\log n)$, the time is $O(n^4\log n+n^2b)$.
\end{proof}

Algorithm \Windows\ relies on two properties of our model.  A closed
exploration needs the ports by which the agent enters, and the frame tables
are found by exhaustive search.  For agents that start simultaneously, the
algorithm becomes much simpler.

\begin{remark}[Agents that start simultaneously]
\label{rem:simultaneous}
Agents that start simultaneously need neither waiting periods nor closed
explorations.  In stage $i\ge1$, an agent executes the frame for
$B=T=2^i$ that consists of $2T$ free rounds only.  Stage $i$ takes
$2^{i+1}$ rounds in any graph, so both agents start every stage in the same
round.  Consider the first stage with $2^i\ge256\,h(n)(A(n)+b)$.  In this
stage $B\ge n$.  By \cref{lem:frames} with $\tau=0$, the agents meet in
this stage at the latest.  The stages up to this stage take fewer than
$2^{i+2}$ rounds, which is $O(n^4\log n+n^2b)$.
\end{remark}

\begin{remark}[Gathering]
\label{rem:gathering}
Consider a team of agents with distinct labels that start in arbitrary
rounds at distinct nodes, and suppose that agents at the same node can
exchange their states.  Kowalski and Malinowski~\cite[Lemma~2.1]{KM08} turn
a rendezvous algorithm into a gathering algorithm as follows.  Each agent
executes the rendezvous algorithm with its label.  When agents meet, all of
them follow from then on the agent with the smallest label among them, and
this agent continues its own execution as if no meeting had occurred.
With Algorithm \Windows\ as the rendezvous algorithm, the team gathers at one node within
$O(n^4\log n+n^2b)$ rounds after the start of the last agent, where $b$ is
the length of the smallest label.  Indeed, call an agent a \emph{leader} if
it follows no other agent.  An agent that follows another agent never
becomes a leader again.  Hence every leader has executed Algorithm
\Windows\ with its own label since its start, as if it were alone.  The
agent with the smallest label is always a leader.  By \cref{thm:upper}, it
meets every other agent that is still a leader within
$O(n^4\log n+n^2b)$ rounds after the start of the later of the two agents, and after
this meeting the other agent is no longer a leader.  Hence, after this
time, the agent with the smallest label is the only leader, and all agents
are at its node.  The agents do not detect that the gathering is complete.
\end{remark}

\section{The logarithmic factor}
\label{sec:log}

For labels of length $O(n^2)$, our bounds are $\Omega(n^4)$ and
$O(n^4\log n)$.  We could not close this gap.  In this section, we show two
facts about it.  First, the factor $\log n$ is absent on all covers of a
known base.  Hence a larger lower bound cannot be proved on the graphs used
in \cref{sec:quartic}.  Second, an algorithm with time $O(n^4)$ would yield
short words that traverse all edges of certain dense graphs.  No such
words are known.

\subsection{Covers of a known base}

\begin{proposition}
\label{prop:known-base}
Let $B$ be a connected port-labelled graph of size $q$, and let
$k\ge2$.  Suppose that both agents know $B$ and $k$.  There exists an
algorithm that guarantees rendezvous within $O(n^4+n^2b)$ rounds after the
start of the later agent, in every connected $k$-sheet cover of $B$ and for
any delay.  Here $n=kq$ is the size of the cover.
\end{proposition}
\begin{proof}
A $k$-sheet cover of $B$ is given by one permutation of the $k$ sheets for
each edge of $B$.  The graph $B$ has at most $q^2/2$ edges.  Hence the
number $F$ of the covers is at most $(k!)^{q^2/2}$, and
\[
 \log F\le\frac{q^2}2\,k\log k=\frac{\log k}{2k}\,n^2\le n^2 .
\]
The proposition follows from \cref{prop:cyclic} with $N=n$.
\end{proof}

The lower bound of \cref{cor:quartic-fourcover} is proved on four-sheet
covers of a clique that the agents may know.  By \cref{prop:known-base}, a
lower bound that is proved on the covers of a known base cannot exceed the
order $n^4+n^2b$, whatever the number of sheets.  The graphs $G(w_1,w_2)$
of \cref{thm:quartic} are not always covers of a clique.  However, for a
fixed $n$ they form a family that agents that know $n$ can compute.  A graph
of this family is given by two signings, by the two removed edges and, for
odd $n$, by the ports of the new node.  Hence $\log F=O(n^2)$ for this
family, and by \cref{prop:cyclic} the agents meet on it within
$O(n^4+n^2b)$ rounds.  A lower bound of order
$n^4\log n$, if it is true, needs a family of graphs with $\log F$ of order
$n^2\log n$.

\subsection{Words that traverse all edges}

The upper bound $O(n^4\log n)$ comes from the estimate
$\log F=O(n^2\log n)$ for the family of all graphs.  We now show that most
of this estimate is already needed for graphs that give no information to
an agent.

Let $d\ge1$.  An \emph{involution} of $\{1,\dots,d\}$ is a permutation
$\pi$ such that $\pi(\pi(p))=p$ for every $p$.  Fix such an involution.  We
say that a $d$-regular port-labelled graph is
\emph{$\pi$-symmetric} if every edge that has port number $p$ at one end
has port number $\pi(p)$ at the other end.  For example, the covers of the
clique $K_q$ with the cyclic port labelling are $\pi$-symmetric for
$\pi(p)=q-p$.  As in \cref{lem:words}, an agent executes a fixed word in all
$\pi$-symmetric graphs.  Let $\mathcal S_\pi(n)$ be the family of connected bridgeless
$\pi$-symmetric graphs of size $n$.  We say that a
word is \emph{edge-universal} for $\mathcal S_\pi(n)$ if its walk traverses
every edge, in every graph of $\mathcal S_\pi(n)$ and from every node.  We
denote by $E_\pi(n)$ the minimum length of such a word.  Edge-universal
sequences were introduced by Borodin et al.~\cite{BRT92} for the family of
all $d$-regular graphs of a given size.

\begin{proposition}[Reduction]
\label{prop:reduction}
Suppose that an algorithm guarantees rendezvous within $T$ rounds in all
graphs of size $2n$, for some pair of distinct labels, for agents that
start simultaneously and know the size.  Then $T\ge E_\pi(n)$, for
every $d$ and every involution $\pi$ of $\{1,\dots,d\}$ such that
$\mathcal S_\pi(n)$ is not empty.
\end{proposition}
\begin{proof}
Suppose, for a contradiction, that $T<E_\pi(n)$.  In $\pi$-symmetric graphs
an agent sees the degree $d$ and enters by port $\pi(p)$ after taking port
$p$.  Hence, as in the proof of \cref{thm:quartic}, each agent executes a
fixed word of length $T$ as long as the agents do not meet.  Neither word
is edge-universal.  Thus the walk of the first word from some node $x_1$
of some graph $G_1\in\mathcal S_\pi(n)$ does not traverse some edge
$\{y_1,z_1\}$.  Similarly, the walk of the second word from some node $x_2$
of some graph $G_2\in\mathcal S_\pi(n)$ does not traverse some edge
$\{y_2,z_2\}$.  Remove both edges, and add the edges $\{y_1,y_2\}$ and
$\{z_1,z_2\}$ with the port numbers of the removed edges.  The resulting
graph is simple and has size $2n$.  It is connected, because $G_1$ and
$G_2$ are bridgeless.  Let the agents start at $x_1$ and $x_2$.  Each agent
stays in its own graph and observes what it would observe in $G_1$ or in
$G_2$.  Hence the agents do not meet during the first $T$ rounds.  This
contradiction proves the proposition.
\end{proof}

\Cref{lem:hide} shows that $E_\pi(2q)\ge\binom q2\binom{q-1}2$ for
$\pi(p)=q-p$, and \cref{prop:reduction} applied to this bound gives the
bound on the time in \cref{thm:quartic} for $n$ divisible by $4$.  \Cref{prop:reduction} has two consequences.

The first consequence concerns upper bounds.  A rendezvous algorithm with
time $O(n^4)$ would give edge-universal words of length $O(n^4)$ for every
family $\mathcal S_\pi(n)$, in particular for $d=\Theta(n)$.  The only upper
bound on $E_\pi(n)$ that we know is of order $dn^3\log n$, which is
$n^4\log n$ for $d=\Theta(n)$.  It follows from the argument of Aleliunas et
al.~\cite{AKLLR79}.  A random word traverses a given edge of a $d$-regular graph
within $O(n^2)$ expected steps.  Indeed, on a regular graph the lazy
Metropolis walk is the simple random walk slowed down by the factor $2$.
Hence, by \cref{lem:hit}, the walk of a random word reaches an end $y$ of
the edge within $O(n^2)$ expected steps.  Afterwards it returns to $y$
every $n$ steps in expectation, and at each visit it traverses the edge with
probability $1/d$.  Thus it traverses the edge within $dn+1$ further expected
steps.  Moreover, $\mathcal S_\pi(n)$
has at most $n^{dn}$ graphs.  Hence a random word of length
$O(dn^3\log n)$ is edge-universal for $\mathcal S_\pi(n)$ with positive
probability.  For universal traversal sequences of
$d$-regular graphs, the upper bound $O(dn^3\log n)$ of Kahn et
al.~\cite{KLNS89}, which we quote from~\cite{BRT92}, comes from the same
argument.  We know of no proof that the lower bound $\Omega(d^2n^2)$ of
Borodin et al.~\cite{BRT92} is tight for $d=\Theta(n)$.  Borodin et al.\
themselves point out that, for degrees linear in $n$, their lower bound
and the upper bound $O(n^4\log n)$ differ by a logarithmic factor, and that
a larger lower bound needs a family of graphs richer than the covers of one
base.  Thus, for short labels, an upper bound that closes the gap of our bounds
requires closing a gap of the same kind as the one that Borodin et al.\
left open for universal traversal sequences of dense graphs.

The second consequence concerns lower bounds.  A bound
$E_\pi(n)=\Omega(n^4\log n)$ for some involution with $d=\Theta(n)$ would
show that \cref{thm:upper} is tight.  Such a bound cannot be proved on
the $\pi$-symmetric covers of a fixed base.  Indeed, these covers form a
family of $F$ graphs with $\log F=O(n^2)$.  By the argument of the previous
paragraph, a random word of length $O(n^2\log(Fn^3))=O(n^4)$ traverses
every edge of every graph of this family, from every node, with positive
probability.  A $\pi$-symmetric graph of degree $d$ is described by
$O(dn\log n)$ bits.  A two-sheet cover of
a fixed clique is described by only $\Theta(n^2)$ bits.  A proof of the
larger bound would have to use the remaining freedom.

\begin{remark}[More sheets do not help]
\label{rem:sheets}
A natural attempt is to use covers with $k$ sheets, for a prime $k>2$.  The
integers $0,\dots,k-1$ with addition and multiplication modulo $k$ form a
field.  Fix an order of the two ends of every edge.  Choose first the edge to be
hidden, and a spanning tree of $K_q$ that contains it.  Assign such an
integer $\sg(e)$ to every edge $e$ outside this tree, and let $\sg$ be $0$
on the tree.  Let a traversal of $e$ from its first end to its second end
add $\sg(e)$ to the sheet, modulo
$k$, and let a traversal in the other direction subtract $\sg(e)$.  The
sheet of the agent is then a scalar product $\sg\cdot\chi$, in which the
vector $\chi$ counts the traversals of every edge with the signs of their
directions.  In order to hide the copy of an edge in sheet $1$, the vector $\sg$ of
these $\beta$ integers must be nonzero and must violate one
equation $\sg\cdot\chi=1$ for each traversal of the edge.  If $\chi$ is the
zero vector, the equation has no solution.  Otherwise its solutions
form an affine hyperplane that does not contain the zero
vector.  By the theorem of Jamison~\cite{Jam77} and of Brouwer and
Schrijver~\cite{BS78}, fewer than $\beta(k-1)$ such hyperplanes do not
cover all nonzero vectors.  Hence an edge with fewer than
$\beta(k-1)$ traversals can be hidden.  The number of traversals grows by
the factor $k-1$.  However, each part now has $kq$ nodes, so $n=2kq$.  The
resulting lower bound is $\binom q2\beta(k-1)=\Theta(n^4/k^3)$.  It is
largest for $k=2$, in agreement with \cref{prop:known-base}.
\end{remark}

\section{Conclusion}
\label{sec:conclusion}

We proved that deterministic rendezvous of two labelled agents takes
$\Omega(n^4)$ rounds in the worst case over the graphs of size $n$.  We
also showed that agents that know nothing about the graph and start with an
arbitrary delay meet in time
$O(n^4\log n+n^2b)$, and that the term $n^2b$ is necessary up to the factor
$O(\log n)$ for even $n$, for agents that do not know the other label.
Hence, for labels of length $O(n^2\log n)$, the minimum time of rendezvous
is determined up to the factor $O(\log n)$, and it is the same up to this factor for agents that
know $n$ and start simultaneously.  We conclude with open problems.
\begin{enumerate}
\item A natural open problem is to close the gap between $\Omega(n^4)$ and
$O(n^4\log n)$.  By
\cref{prop:known-base}, the factor $\log n$ is absent on covers of a known
base.  By \cref{prop:reduction}, removing it from the upper bound requires
edge-universal words of length $O(n^4)$ for dense $\pi$-symmetric graphs.
\item Is there an algorithm with time $\widetilde O(n^4+n^2b)$ for agents
that do not learn the ports by which they enter?  Is there one whose local
computations take polynomial time?  Algorithm \Windows\ needs the ports by which
the agent enters for Procedure \Transition\ and for its closed explorations, and our
tables are found by exhaustive search.  For agents that do not learn the
ports by which they enter, the best published bound is
$\widetilde O(n^5b)$~\cite{TZ14}, by an algorithm whose local computations
are not known to take polynomial time.  With polynomial local computations, the best
published time is a polynomial of larger degree~\cite{TZ14}.
\item Close the gap between $\Omega(n^2b/\log n)$ and $O(n^2b)$ in the
dependence of the time on the length of the labels.  By
\cref{rem:labels-limit}, the graphs used for \cref{thm:labels} cannot give
more.  The lower bound is proved only for the sizes $n$ that are divisible
by $2$ or by $3$ (\cref{rem:three}).
\item What is the minimum time of rendezvous when the initial distance is
$4$, $5$ or $6$?  By \cref{rem:why-seven}, a quartic lower bound for these
distances needs a different argument.
\item How does the time of gathering depend on the size $k$ of the team?
By \cref{rem:gathering}, every team gathers within $O(n^4\log n+n^2b)$
rounds.  \Cref{cor:teams} gives $\Omega(n^4/k^4)$ for a team of $k$ agents,
and by~\cite{MMM23} a team of more than $n/2$ agents that know $n$ and
start simultaneously gathers in $O(n^3)$ rounds.
\item Is there an explicit exploration procedure of length
$\widetilde O(n^4)$?  By \cref{cor:treasure} no procedure is shorter than
$(1-o(1))\,n^4/64$, and the procedure of \cref{rem:explore} is found by
exhaustive search.
\item Improve the constant $1/1024$ of \cref{thm:quartic}.  Our construction
uses $4q$ nodes for a base clique of size $q$.  By \cref{rem:hide-tight},
hiding an edge from a longer word is not possible in two-sheet covers of
$K_q$.
\end{enumerate}

\section*{Acknowledgement of the use of AI}

This paper was written with the help of AI models.  The upper bound
(\cref{thm:upper} and Algorithm \Windows) was first proposed by Astra-6.
The lower bound (\cref{thm:quartic}) was found in conversations of the
author with Opus~5.5 and Fable~5.1 over two weeks.  The models also helped
to write the text and to check the proofs.  The author is responsible for
the content of the paper.

\end{document}